%% file: main.tex
\PassOptionsToPackage{expansion=false}{microtype}
\documentclass[sigconf,nonacm]{acmart}

\setcopyright{none}
\selectcolormodel{rgb}

\definecolor{revisionred}{RGB}{0,0,0}
\definecolor{shepherdred}{RGB}{0,0,0}
\newcommand{\shepadd}[1]{\textcolor{shepherdred}{#1}}
\newenvironment{shepherding}{\begingroup\color{shepherdred}}{\endgroup}
\input{generated/metrics.tex}

\newcommand{\revadd}[1]{\textcolor{revisionred}{#1}}
\newenvironment{revision}{\begingroup\color{revisionred}}{\endgroup}
\newcommand{\revcaptionsetup}{%
  \captionsetup{labelfont={color=revisionred},textfont={color=revisionred}}}
\usepackage{tikz}
\usetikzlibrary{arrows.meta,positioning,fit,backgrounds,calc,decorations.pathreplacing}

\usepackage{graphicx}
\usepackage{adjustbox}

\usepackage{float}
\usepackage{dblfloatfix}
\usepackage{array}
\usepackage{booktabs}
\usepackage{longtable}
\usepackage{comment}
\usepackage{needspace}

\newcommand{\jag}[1]{}
\newtheorem{innercustomgeneric}{\customgenericname}
\providecommand{\customgenericname}{}
\newcommand{\newcustomtheorem}[2]{%
  \newenvironment{#1}[1]
  {%
   \renewcommand\customgenericname{#2}%
   \renewcommand\theinnercustomgeneric{##1}%
   \innercustomgeneric
  }
  {\endinnercustomgeneric}
}

\newcustomtheorem{customclaim}{Claim}
\newcustomtheorem{customthm}{Theorem}
\newcustomtheorem{customlemma}{Lemma}
\newcustomtheorem{customdefinition}{Definition}
\newcustomtheorem{customexample}{Example}
\newcustomtheorem{customproblem}{Algorithmic Problem}

\input{authors.tex}

\begin{document}
\setlength{\emergencystretch}{2.5em}

\title{ConStruM: A Structure-Guided LLM Framework for Context-Aware Schema Matching}



\begin{abstract}
Column matching is a central task in reconciling schemas for data integration. Column names and descriptions are valuable for this task. LLMs can leverage such natural-language schema metadata.
However, in many datasets, correct matching requires \emph{additional evidence beyond the column itself}.
Because \revadd{providing complete schema metadata remains impractical for
most LLMs and can obscure the relevant evidence even when it fits}, the core challenge
becomes to select and organize the most useful contextual information.

We present \textbf{ConStruM}, a structure-guided framework for \emph{budgeted evidence packing} in schema matching.
ConStruM constructs a lightweight, reusable structure in which, at query time, it assembles a small \emph{context pack} emphasizing the most discriminative evidence.
ConStruM is designed as an \emph{add-on}: given a shortlist of candidate targets produced by an upstream matcher, it augments the matcher’s final LLM prompt with structured, query-specific evidence so that the final selection is better grounded.
For this purpose, we develop a \emph{context tree} for budgeted multi-level
context retrieval and a \emph{global similarity hypergraph}
\revadd{for grouping and differentiating confusable columns}.
\begin{revision}
Across HRS-B and PSID-B, ConStruM \shepadd{achieves a mean combined accuracy of
\SurveyFullCombined{}, compared with \SurveyNoContextCombined{} for the
no-context LLM}, and ranks first among all \shepadd{eleven}
evaluated methods on both benchmarks.  On MIMIC--OMOP and Synthea--OMOP,
attaching it to our Matchmaker reconstruction produces no statistically
significant accuracy change.  Together, these results show that ConStruM can
deliver large gains when matching depends on context while remaining
compatible with a strong matcher in the other tested settings.
\end{revision}
\end{abstract}

\maketitle

\begin{figure}[!t]
  \centering
  \Description{A four-rectangle schematic. Source (MIMIC-III): query CHARTTIME with a dashed context note from another column (STORETIME). Target: two candidate timestamps. Two identical arrows indicate uncertainty between candidates.}
  \input{figures/intro_time_example_tikz.tex}
  \caption{A context-critical schema-matching example from MIMIC-III: \texttt{CHARTTIME} is ambiguous between “time observed” and “time recorded/entered” targets; the sibling column \texttt{STORETIME} provides the clue to choose the right meaning.}
  \label{fig:intro_time_example}
\end{figure}

\input{figures/intro_strategy_comparison_float.tex}

\section{Introduction}

Schema matching---aligning semantically equivalent columns across datasets---is a foundational problem in data integration~\cite{rahm2001survey,doan2005semanticintegration,bellahsene2011schemamatchingmapping}.
When done well, it unlocks reliable analytics, migration, and integration across widely varied data sources.
Large language models (LLMs) are an appealing backbone for this task because they can leverage natural-language schema metadata~\cite{parciak2024schema, sheetrit2024rematch, ma2025knowledge, seedat2024matchmaker, xu2024kcmf}.
Yet, in many real datasets, semantic correctness depends on \emph{evidence beyond the column itself}: where the column sits in the schema, what neighbors co-occur with it, lightweight relations (e.g., cross-field dependencies or simple cross-references), and how it contrasts with other plausible candidates.

\noindent\textbf{Motivating example.}
Consider the \texttt{CHARTEVENTS} table in MIMIC-III~\cite{johnson2016mimiciii}, a widely used critical care dataset.
Two time fields are easy to confuse from names alone: \texttt{CHARTTIME} (when an observation was made) and \texttt{STORETIME} (when it was entered).
In the documentation used by our benchmark, \texttt{CHARTTIME} is described generically as ``records the time at which an observation was made'', while its sibling \texttt{STORETIME} provides the discriminative clue that it ``records the time at which an observation was manually input or manually validated by a member of the clinical staff.''
Now suppose an upstream matcher proposes two plausible targets: \texttt{observation\_time} (time observed) and \texttt{recorded\_time} (time recorded/entered).
If the model only sees \texttt{CHARTTIME}, both candidates can look plausible; including \texttt{STORETIME} provides the missing context needed to disambiguate them (Figure~\ref{fig:intro_time_example}).

This example is not an isolated corner case.
In many real schemas, column semantics cannot be determined in isolation.
The discriminative evidence needed for correct matching often lies outside the column itself and may be distributed across the schema.

A natural response is to expose the LLM to more context by providing the full source and target schemas in the prompt and asking it to reason over the complete picture~\cite{parciak2024schema}.
However, this \emph{global prompting} strategy breaks down in practice.
Realistic schemas can contain hundreds or thousands of attributes, making
full-schema prompts infeasible under realistic context budgets
\revadd{for most widely used LLMs. A few recent long-context models can accept
prompts at this scale}.
\revadd{Even then, fitting the schema into the context window does not ensure
that the model will use the relevant evidence:} LLMs do not always reliably
use the \emph{right} information when it is buried inside large
prompts~\cite{liu2024lost,bai2024longbench}, and schema-matching experiments
similarly find that increasing prompt scope is not consistently
beneficial~\cite{parciak2024schema}.

As a result, most existing systems adopt a pragmatic alternative: \emph{local prompting}.
While these systems differ in their overall pipelines and auxiliary mechanisms~\cite{seedat2024matchmaker,sheetrit2024rematch,ma2025knowledge,xu2024kcmf}, the LLM is typically asked to reason over a small number of columns at a time, using column names and descriptions and, in some cases, additional table-level text such as the table description.
In settings like the motivating example above, where correct interpretation depends on broader schema context, this formulation forces the model to decide without seeing the specific outside-column cues that resolve the ambiguity.

\noindent\textbf{Our perspective: context-critical schema matching as evidence packing.}
In the subset of schema matching problems where beyond-column evidence is essential, the goal is \shepadd{to \emph{select and organize}} a compact set of evidence that is maximally useful for disambiguation.
We call this compact, query-specific bundle a \emph{context pack}.

We develop \textbf{ConStruM}, a structure-guided framework for constructing context packs for schema matching.
ConStruM constructs explicit intermediate structure over the schema, and uses it to assemble query-specific context packs that are supplied to the final matching decision.
Figure~\ref{fig:local-global-graph} situates ConStruM among common prompting strategies for LLM-based schema matching---\emph{local}, \emph{global}, and our \emph{structure-guided} approach---and illustrates how ConStruM bridges the gap by assembling a compact, query-specific context pack instead of pasting the full schema.

\noindent\textbf{Add-on workflow (high level).}
ConStruM is not an end-to-end schema matcher.
Instead, it is best viewed as an add-on to the local pipeline described above, supplying additional structured context when column-level evidence alone is insufficient.
Importantly, this form of contextual augmentation fits naturally with existing matching methods:
it introduces additional evidence at the point of comparison, without interfering with how those methods operate.
ConStruM performs offline preprocessing to construct intermediate data structures, including a context tree over schema elements and a global similarity hypergraph.
At use time, an upstream method proposes a set of plausible target columns for a source column. ConStruM then queries these data structures to assemble a context pack,
consisting of multi-level schema context and explicit contrast among confusable columns.
This context pack is appended to the matching prompt, enabling more reliable disambiguation.

Of course, selecting the appropriate query-specific context is easier said than done. A tempting baseline is to treat the \emph{entire table} (or documentation section) as context, but this can be either too large to be useful or too coarse to resolve ambiguity among confusable candidates scattered across a schema.
In practice, the most useful evidence is \emph{multi-level}: provide a coarse-to-fine view that pairs a small amount of local evidence with progressively higher-level summaries. Finally, disambiguation is inherently \emph{comparative}: when several candidates look similar, the prompt should encourage explicit contrast among them rather than evaluating each candidate in isolation.

We instantiate ConStruM with two lightweight data structures that directly target these needs:
(i) a \emph{context tree} that encodes hierarchical schema neighborhoods and supports budgeted retrieval of a compact multi-level context pack for any column, and
(ii) a \emph{global similarity hypergraph} that surfaces groups of highly similar columns (on both source and target sides) so the prompt can present confusable options together, augmented with concise group-wise differentiation cues (computed on demand or precomputed offline).

\begin{revision}
We evaluate ConStruM on two groups of benchmarks.  \textbf{HRS-B} and
\textbf{PSID-B} are context-stress benchmarks built from 
independently maintained longitudinal surveys.  They contain
confusable fields whose interpretation depends on broader questionnaire
context.  \textbf{MIMIC--OMOP}~\cite{sheetrit2024rematch} and
\textbf{Synthea--OMOP} evaluate the same add-on idea on general schema
matching.  ConStruM ranks first among the \shepadd{eleven} methods evaluated on both
context-stress benchmarks.  On the two general schema-matching benchmarks,
adding ConStruM to our Matchmaker reconstruction produces no statistically
significant accuracy change.
\end{revision}

\Needspace{4\baselineskip}
\paragraph{Contributions.}
\begin{itemize}
  \item \textbf{Perspective + add-on framework.} We study context-critical schema matching as an \emph{evidence packing} problem and propose \textbf{ConStruM}, a structure-guided context module that can be attached to existing LLM-based matchers. ConStruM builds compact context packs \revadd{using a context tree and a global similarity hypergraph}.
  \item \textbf{\revadd{Benchmarks}.} We introduce \textbf{HRS-B} \revadd{and \textbf{PSID-B},
  context-stress benchmarks built from 
  the Health and
  Retirement Study~\cite{HRS} and Panel Study of Income Dynamics~\cite{psid}
  (Section~\ref{sec:hrs}). Both deliberately select field pairs that are
  difficult to match from field text alone and require understanding their surrounding schema context}.
  \item \begin{revision}\textbf{Evaluation.} We compare \shepadd{eleven} methods on both
  context-stress benchmarks and evaluate the Matchmaker add-on on two
  general schema-matching benchmarks. \shepadd{The three-run comparison ranks
  ConStruM first on both surveys}; ablations and supporting controls explain the
  contribution, scope, and measured cost of structured evidence.\end{revision}
\end{itemize}

\section{Problem Setting and Preliminaries}
\label{sec:preliminaries}
We study \emph{column-level schema matching} between a source schema
\revadd{\(S=\{s_1,\ldots,s_m\}\)} and a target schema
\revadd{\(T=\{t_1,\ldots,t_n\}\)}.  At use time, the input is a single source
column \(s\in S\), and the task is to select a semantically equivalent target
column \(t\in T\).  Equivalently, over all source columns this yields a mapping
\(M:S\rightarrow T\); we focus on a \emph{forced-choice selection} setting
where every \(s\) selects exactly one \(t\).  We do not require \(M\) to be
globally one-to-one unless enforced by a downstream application.

\begin{revision}
This is the conventional schema-matching notion of a correspondence between
schema elements~\cite{rahm2001survey,doan2005semanticintegration,bellahsene2011schemamatchingmapping}:
the two columns represent the same attribute or field role.  Our scope is
specifically single-target matching.
Each decision produces one source--target pair, although different sources may
select the same target.  Calibrated no-match decisions and one-to-many or
many-to-many transformations are important extensions, but fall outside this
setting.
\end{revision}

An alternative formulation assigns a compatibility score \(f(s,t)\) to each
pair and then selects a match by \(\arg\max_{t\in T} f(s,t)\).  In this paper we
directly define and evaluate the selection problem \(M:S\rightarrow T\),
consistent with our benchmarks and evaluation protocol.  Accordingly, we
\revadd{report exact-match accuracy at the source-decision level}.

\paragraph{Example.}
Consider:
\begin{itemize}
  \item \textbf{Source columns:} \texttt{name}, \texttt{zip\_code}.
  \item \textbf{Target columns:} \texttt{customer\_name}, \texttt{postal\_code}.
\end{itemize}
After preprocessing \(S\) and \(T\) (e.g., indexing/embedding \(T\)), a query might take \(s=\texttt{zip\_code}\) as input and output \(t=\texttt{postal\_code}\).

\section{Method Overview}
\label{sec:artifacts}

\subsection{Context as Add-on Evidence}
\label{sec:overview-addon}
Many LLM-based schema matchers, despite differing in upstream retrieval or multi-stage reasoning, share a common \emph{decision-time pattern}: given a query/source column \(s\) and a small shortlist of plausible target candidates \(C_0\), the final step is a single LLM call that selects one target from \(C_0\), such as ReMatch~\cite{sheetrit2024rematch} and Matchmaker~\cite{seedat2024matchmaker}.
In this final step, the LLM functions as an \emph{evidence aggregator}: it commits to a choice by synthesizing the evidence presented in the prompt.

If the final choice is evidence-driven, then any information that is useful for disambiguation can be supplied as \emph{additional decision-time evidence}---without modifying how \(C_0\) was produced.
Contextual augmentation is therefore inherently add-on: it strengthens the final decision by enriching the evidence available for \(s\) and candidates in \(C_0\), while leaving upstream candidate generation untouched.

\subsection{When to Compute Schema Context}
\label{sec:overview-offline-online}
Once context is treated as decision-time evidence, it becomes unnecessary---and often undesirable---to compute all such evidence only at query time, after \((s, C_0)\) is observed.
Schema context is a schema-level resource: it is structural and reused across repeated match queries on the same schemas.
This naturally leads to a separation between offline preprocessing and online
query-time instantiation.

\paragraph{Offline preprocessing.}
Offline preprocessing should compile reusable schema context into intermediate data structures that can be amortized across many queries, potentially using LLM calls for summarization or abstraction.

\paragraph{Online instantiation.}
Online instantiation should \emph{address} these offline data structures with the specific \((s, C_0)\) at hand, instantiating only a compact subset that fits the prompt budget---a \emph{context pack} tailored to the decision among candidates.

\subsection{The Intermediate Data Structure}
\label{sec:overview-structure}
The remaining question is what properties these intermediate data structures
must satisfy so that online instantiation produces useful evidence
under a fixed prompt budget.
Two recurring failure modes can be understood as limitations of naive context organization, and they translate directly into representational requirements.

\noindent\textbf{Context must be selected and organized under a prompt budget.}
Because the context pack must fit in a bounded prompt, the intermediate data
structures should support hierarchical, coarse-to-fine scope instantiation:
broad schema context can be summarized compactly, and finer detail can be
allocated only where it is most diagnostic for the decision.

\noindent\textbf{Confusable alternatives require contrast.}
When alternatives are close, similarity-based reasoning becomes intrinsically underdetermined unless the decision is framed as an explicit comparison that exposes the axes of difference.
This applies both within \(C_0\) and around \(s\), whose intended meaning may only be recoverable by contrast with closely related alternatives, including near-duplicates in some schemas.
The intermediate data structures should therefore surface \emph{sets} of
confusable alternatives around \(s\) and within \(C_0\), and provide compact
contrast cues so the decision-time prompt can place close options side-by-side.

\subsection{The ConStruM Framework}
\label{sec:overview-instantiation}
ConStruM organizes schema context into two complementary data structures built during offline preprocessing.
A \emph{context tree} supports hierarchical, coarse-to-fine context organization (Section~\ref{sec:tree}), while a \emph{global similarity hypergraph} captures groups of confusable alternatives to enable contrastive reasoning (Section~\ref{sec:group}).
At match time, these offline-built structures are used to assemble a compact \emph{context pack} for \((s, C_0)\) and, when enabled, an expanded working set derived from \(C_0\); this evidence is appended to the final decision prompt (Section~\ref{sec:pipeline}).
Figure~\ref{fig:method-overview-flowchart} summarizes this add-on interface and the offline/online split.

\begin{figure}[t]
  \centering
  \Description{A general LLM-based matcher pipeline uses upstream stages to produce a shortlist (C0) of target candidates for a query/source column (s), then makes a final LLM decision over a small working set. ConStruM is a dashed add-on: it builds a context tree and global similarity hypergraph offline, and online assembles a context pack and query-local confusable groups, optionally expands the working set, and appends multi-level context and contrast cues without changing upstream candidate generation.}
  \input{figures/method_overview_flowchart_tikz.tex}
  \color{revisionred}
  \revcaptionsetup
  \caption{Method overview: ConStruM as add-on \emph{evidence packing}.
  Upstream stages produce a shortlist \(C_0\) of target candidates for a
  query/source column \(s\). ConStruM builds a context tree and global
  similarity hypergraph offline; at match time, it uses them to assemble a
  context pack and query-local confusable groups for \((s,C_0)\). It may
  expand the small working set and append multi-level context and contrast
  cues to the final decision prompt. The LLM then selects the match without
  changing upstream candidate generation.}
  \label{fig:method-overview-flowchart}
\end{figure}

\section{Tree-Based Context Module}
\label{sec:tree}

\subsection{Motivation: multi-level context via hierarchical retrieval}
In realistic schemas, a column’s meaning often depends on both its immediate neighborhood and broader schema organization. Simply appending large neighborhoods to the prompt is often ineffective: it can exceed practical budgets and can introduce irrelevant noise that obscures the most discriminative cues.
\revadd{Our goal is to produce a \emph{query-specific, structured} view of
context that fits a prompt budget and is easy for the model to navigate.}

\paragraph{Intuition (analogy).}
This is similar to debugging an unfamiliar query in a large database system: looking only at a single operator or predicate is rarely enough, but dumping the entire schema and workload context is overwhelming.
Instead, practitioners rely on a coarse-to-fine view---high-level scope (which subsystem/table family is involved) plus local details (relevant attributes, keys, and constraints) that are most diagnostic for the question at hand.
Schema matching under a prompt budget has the same shape: we want to provide a small amount of information at multiple granularities so the model can infer scope and meaning without being flooded by irrelevant details.

Under a fixed prompt budget, we must therefore decide \emph{how much} context to include. Too little context can omit necessary scope cues; too much context wastes budget and buries the relevant evidence. Our key intuition is that useful context is inherently \emph{multi-resolution}: broad scope can be summarized coarsely, while local neighborhoods can be described more precisely, keeping the amount of information at each level comparable.

\begin{figure}[t!]
\centering
\Description{A schematic context tree for schema matching. Each column is a leaf, and each internal node summarizes a broader group of columns formed by merging lower-level ones. The tree depth varies by dataset; for readability we show only a few levels (three here) and one highlighted lineage from the query column leaf up to the root.}
\includegraphics[width=\columnwidth]{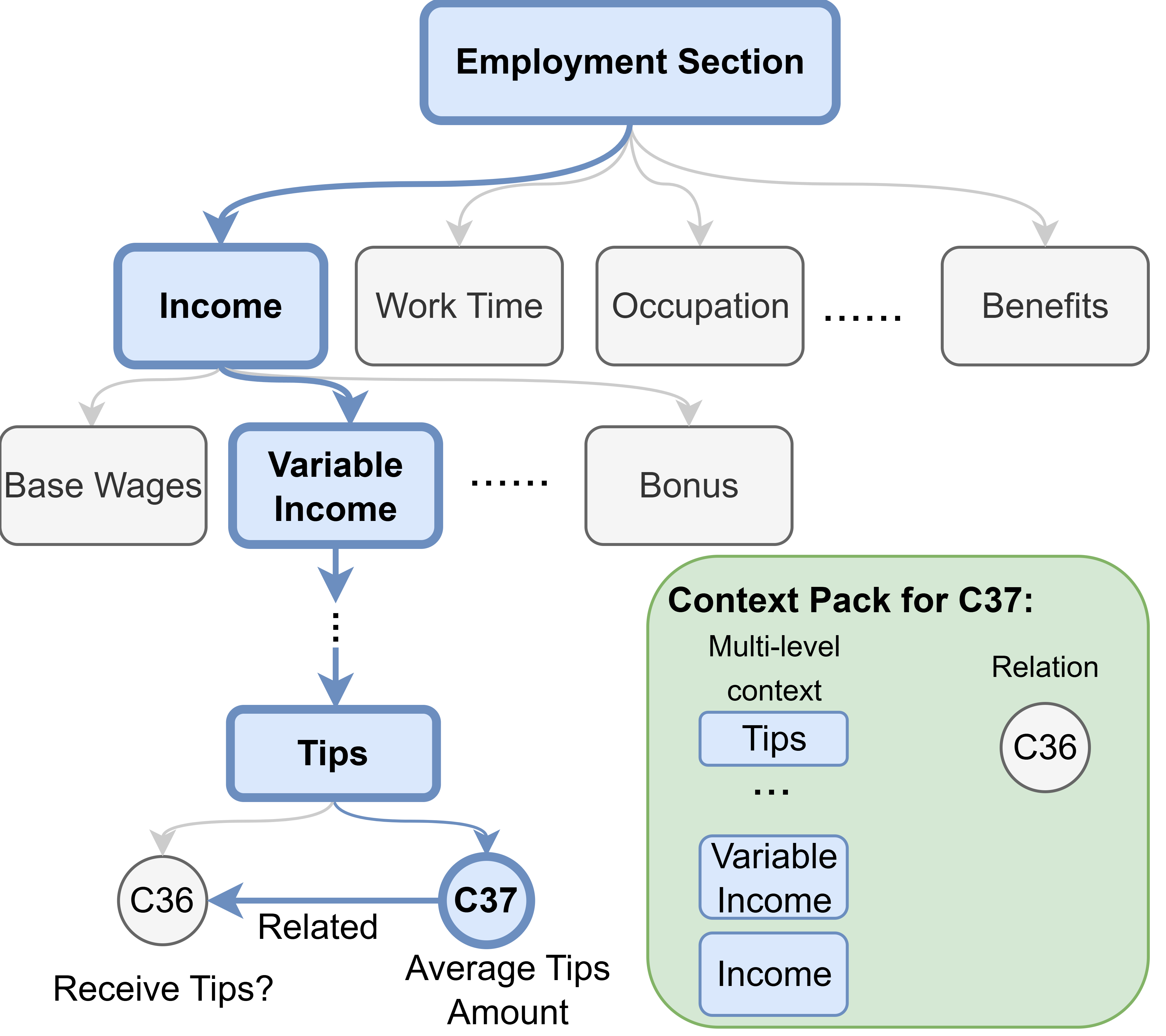}
\caption{Multi-level context retrieval with a context tree. During tree construction, ConStruM may annotate local relation snippets when useful local structural relations are present. For a query column, it then retrieves the column-to-root lineage of increasingly coarse summaries and, when available, relevant local relation snippets. Together they provide local cues and broader schema context under a fixed prompt budget.}
\label{fig:context-tree-overview}
\end{figure}

We realize this idea with a \emph{context tree}, a hierarchical index over the
schema.  \revadd{RAPTOR~\cite{sarthi2024raptor} inspires the broad idea of
organizing evidence at several resolutions.  ConStruM's tree is
schema-specific: it preserves field identities and native table or
documentation scope, and retrieves the relevant lineages for a source field
and its candidate targets.}
Each \emph{column} is treated as a leaf in this tree.
Internal nodes summarize progressively broader groups of columns formed by merging lower-level ones.
During tree construction, ConStruM can also annotate optional \emph{local relation snippets} when useful structural relations are present.
At query time, given a query column, ConStruM follows the path from that column leaf up to the root, collecting summaries along the way, and retrieves relevant local relation snippets when they are available.
This \emph{column-to-root lineage}, together with optional local relation snippets, gives the LLM a coarse-to-fine view of relevant context under a fixed prompt budget (Figure~\ref{fig:context-tree-overview}).
This lineage-plus-relation evidence is the backbone of the compact \emph{context pack} that we feed to the downstream matching decision (Section~\ref{sec:pipeline}).

\subsection{Construction: turning a database into a context tree}
\label{sec:tree-construction}
The immediate obstacle is that a schema is not a single document---it is a collection of \emph{tables}, and most of the strongest locality signals live within tables.
At the same time, treating each table in isolation is too narrow: related tables should contribute shared, higher-level scope.

We therefore construct one database-wide context tree (Figure~\ref{fig:context-tree-overview}) in two stages.
First, we turn \emph{each table} into a small tree whose leaves are columns and whose internal nodes summarize progressively broader within-table regions.
When the schema provides useful local structure, we preserve it; otherwise we fall back to a stable grouping.
Second, we organize tables themselves by clustering the resulting per-table trees into a database-level hierarchy, repeatedly merging the most related trees under a new parent until a single root remains.

This ``tree of trees'' construction naturally handles uneven table sizes: small tables yield shallow subtrees, while very wide tables require deeper, multi-pass construction (detailed next) before they can be integrated at the database level.

\subsection{Step 1: turn each table into a tree}
\label{sec:within-table-construction}
Each table becomes its own small subtree.
At the bottom are the column leaves; above them, internal nodes are natural-language summaries of progressively broader within-table scope.
We preserve available local structure when it is useful; otherwise we use a stable grouping for batching.
\emph{Small} tables are straightforward: a table root summary plus one leaf per column.
The harder case is \emph{wide} tables, which may not fit into one LLM prompt; for these we build a deeper hierarchy via recursive, coarse-to-fine chunking.

\paragraph{Wide tables: recursive chunking into a multi-level tree.}
When a table is wide, our goal is to build a \emph{coarse-to-fine} within-table hierarchy that can later provide both local neighborhood cues and higher-level scope.
Since wide tables may exceed an LLM context window, we build the within-table tree in four stages.
Stages 1--2 build a budget-feasible \emph{global picture} of the table (from windowed batches and sparse samples); Stage 3 converts this evidence into a compact \emph{grouping plan}; and Stage 4 (optional) refines region boundaries when the table provides a meaningful native sequence.
We then recurse on any resulting group/span that is still too large, yielding multiple levels of increasingly coarse within-table summaries.

\noindent\textbf{Stage 1: windowed batch summaries.}
We scan the columns in prompt-sized windows and ask the LLM to write a short summary/theme for each window (schema-native windows when a meaningful sequence exists; fixed-size batches otherwise).

\noindent\textbf{Stage 2: global theme and flow inference.}
In parallel, we obtain a complementary global view from sparse table-wide evidence: using evenly spaced (or random) samples, an LLM infers the overall theme and, when meaningful, a coarse topical progression.
Together, Stages 1--2 provide a compressed, table-level view under input-length limits, which helps stabilize labels and reduces drift across local windows.

\noindent\textbf{Stage 3: unified conceptual map.}
We synthesize the window summaries and sampled evidence into a \emph{canonical} conceptual map: a compact set of semantic groups with standardized labels.
Groups follow schema-native regions when available; otherwise they are semantic sets of columns (with an optional induced presentation sequence).

\noindent\textbf{Stage 4 (optional): boundary refinement and alignment.}
When the table has a meaningful native sequence, we scan it again in prompt-sized windows and align each window to the conceptual map, refining where a new region begins.
\revadd{Continuity constraints suppress short, oscillating fragments.}

\noindent\textbf{Recursion and stopping.}
After Stage 3 (and Stage 4 when applicable), we recurse on any resulting group/span that still exceeds a \revadd{requested} lowest-level group budget \(B\) (\revadd{the target number of} columns summarized together before attaching individual column leaves beneath that group); \(B\) \revadd{guides the tree height as a stopping target rather than a hard bound: \shepadd{recursion stops} when a split makes no progress.  Match-time serialization is bounded separately.}

In practice, this procedure yields stable regions and supports multi-level recursion even for very wide tables.

\subsection{Step 2: cluster per-table trees into one database tree}
After Step 1, we have a collection of per-table trees (deep ones for wide tables, shallow ones for small tables).
The most direct way to make them ``one structure'' is to introduce a single super-root and attach every table root as its child.
However, that construction is nearly flat for real databases: the super-root has many children and provides little reusable, intermediate context.
At match time, it forces an awkward choice between staying within one table (too narrow when relevant evidence is spread across related tables) and pulling in many unrelated tables (too broad to be helpful).

Step 2 builds a database-level hierarchy that creates those missing intermediate layers.
Each per-table tree is represented by a compact text summary at its root (for wide tables, this is the top-level within-table summary; for small tables, it is the table description/summary node).
We embed these root summaries and then perform hierarchical clustering to group semantically related tables.
The database tree is built bottom-up by repeatedly merging a small set of the most related trees under a new parent, and asking an LLM to write a short summary for that parent.
As merging repeats, internal nodes become increasingly broad ``modules'' of tables, yielding a multi-level path from any column to a single database root.

\revadd{Average-link cosine distance determines the merge order.  On the two
multi-table general benchmarks, MIMIC--OMOP and Synthea--OMOP, we retain the
complete hierarchy through its root.  In HRS-B and PSID-B, each survey wave is
one large table, so its within-table tree already covers the complete wave.}

\paragraph{Resulting database tree.}
The output is a single connected tree whose leaves are columns.
\shepadd{A column's lineage runs from its leaf through within-table summaries
to the table root. For a multi-table schema, it then follows clustering
summaries of increasingly broad table groups to the database root
(Figure~\ref{fig:context-tree-overview}).}

\subsection{Local relation annotations (optional)}
The tree already captures containment, and parent summaries explain the shared topic of a region.
At match time, however, what matters is sometimes not only ``what is this region about?'' but also ``how do these nearby parts relate?''
One field may qualify the applicability of another, provide an anchor or identifier used by nearby attributes, supply a unit or type that changes interpretation, or participate in a repeated or parallel local pattern.
Parent summaries are meant to stay compact, so they rarely enumerate such internal links explicitly.
Therefore, during tree construction, ConStruM can optionally infer a small number of short \emph{local relation snippets} that capture these decision-relevant relations.
These snippets are stored with the local region and can be retrieved later as additional decision-time evidence.

\subsection{Context retrieval: building a compact context pack}
Given a query column (and likewise for each candidate), ConStruM retrieves a compact \emph{context pack} that fits a fixed prompt budget.
The pack is a single text block that combines (i) a \textbf{column-to-root lineage} of summaries (Figure~\ref{fig:context-tree-overview}) and (ii) optional \textbf{local relation snippets} when they are informative.
The lineage carries both local cues and progressively broader scope, while the relation snippets capture structural links that are easy to miss in a pure containment summary.

\section{Similarity-Group Differentiation Module}
\label{sec:group}

\subsection{Motivation: helping the LLM distinguish confusable alternatives}
Real schemas can contain \emph{confusable clusters} of semantically close columns.
These clusters can appear on the target side (several candidates remain plausible matches for the same source column) and also on the source side, where a query column’s intended meaning is only clear relative to nearby confusable alternatives.
When there is a clear semantic winner, an LLM can often select the closest match from a shortlist.
But once the shortlist contains a tight cluster, an LLM-only ``pick the most similar one'' decision can easily fail, because names/descriptions do not make the discriminative axis salient and multiple candidates look equally reasonable.
The difficulty is therefore not finding related candidates---it is making a \emph{reliable} choice \emph{within} a confusable cluster under a fixed prompt budget.

\begin{revision}
Battleship organizes tuple-pair representations as a graph and uses their
local structure to focus a limited labeling budget on informative examples in
entity matching~\cite{genossar2023battleship}. Inspired by this high-level
idea, ConStruM uses neighborhoods of schema fields to identify confusable
alternatives and places their distinctions together in one decision prompt.
\end{revision}

For instance, consider several semantically close time fields whose descriptions are short and partially overlapping.
Their names may differ (e.g., \texttt{event\_time} vs.\ \texttt{ingest\_time}).
A retriever (e.g., embeddings) will surface them together, but selecting the right one becomes reliable only when the candidates are placed side-by-side and the prompt forces an explicit contrast about \emph{what stage of the workflow each timestamp refers to}.
ConStruM therefore adds an explicit \emph{grouped differentiation} step: it groups confusable candidates and produces short, group-wise contrastive cues so the final decision is made in direct comparison rather than by independent scoring.

\paragraph{Side benefit: candidate expansion.}
The same grouping structure can also improve candidate selection: starting from an initial shortlist, ConStruM can expand from the strongest retrieved candidates to pull in highly similar neighbors, reducing the chance that hard-to-distinguish alternatives are excluded before the final decision step.

\subsection{Global similarity hypergraph construction (offline)}
To reuse confusable-group structure across many queries, we build a global grouping artifact over each schema (source and target).
The idea is to start from an embedding similarity signal, connect columns that are sufficiently close, and then read off confusable groups from the resulting connectivity.
ConStruM represents these groups as a \emph{global} similarity hypergraph.

\begin{figure}[t]
\centering
\Description{A set-system illustration of a schema similarity hypergraph. Each vertex represents a schema column, and each colored hyperedge encloses a group of two or more confusable columns; hyperedges may overlap.}
\includegraphics[width=0.94\columnwidth]{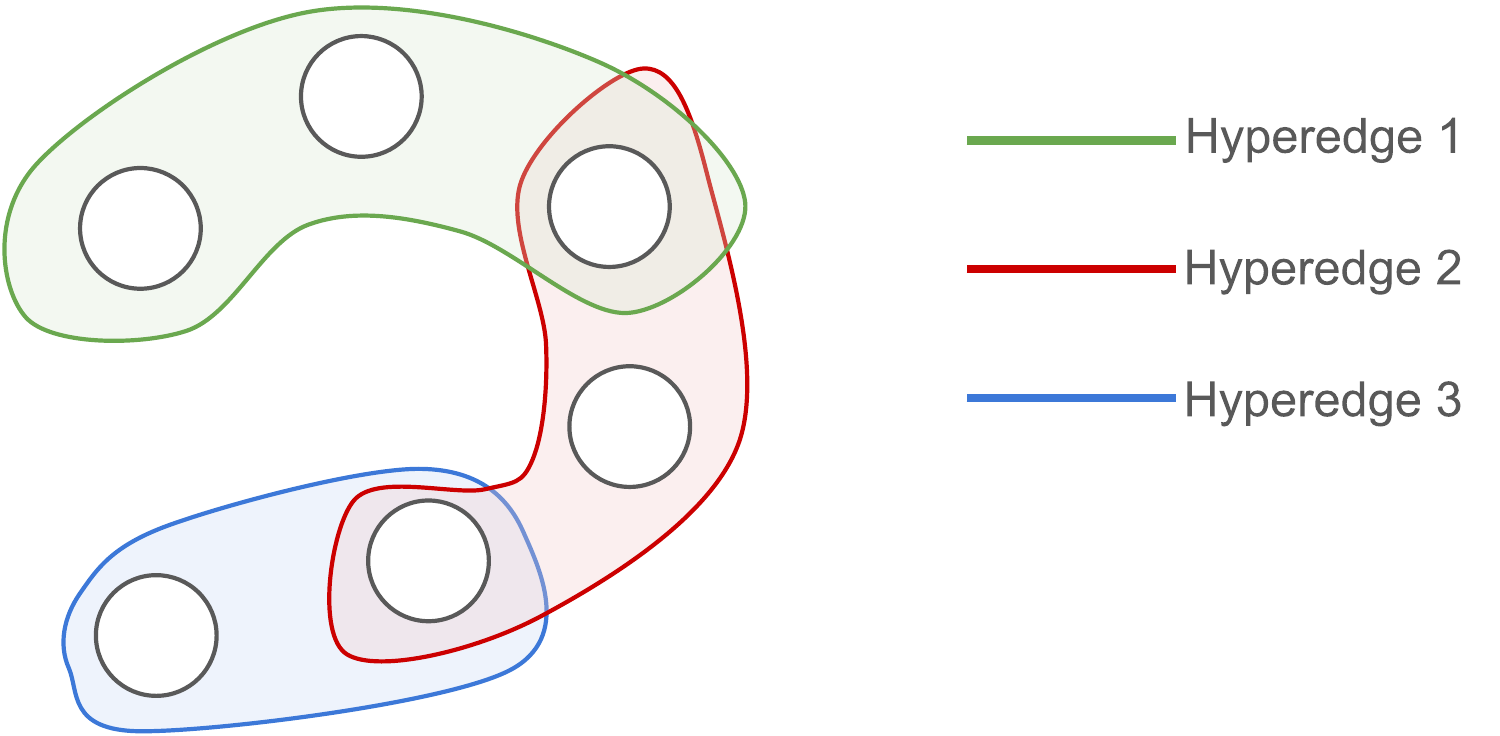}
\caption{A hypergraph differs from an ordinary graph in that each hyperedge can connect multiple vertices rather than just two. Here, each vertex represents a column, and each colored region represents one hyperedge, i.e., one confusable group of columns. Overlap indicates that a column can belong to multiple confusable groups.}
\label{fig:hypergraph-set-system}
\end{figure}

Here a \emph{hypergraph} can be understood simply as a \emph{set system} \((V,\mathcal{E})\): \(V\) is the set of columns in a schema and each hyperedge \(e\in\mathcal{E}\) is a \emph{set} of columns representing one confusable group.
Equivalently, the same information can be represented as a bipartite membership graph between columns and groups.
This representation \shepadd{surfaces} the \emph{entire \revadd{similarity-connected} set} of alternatives that should be contrasted (and, when useful, considered for candidate expansion).
\revadd{A hyperedge is therefore not a topical community to summarize; it is a
set of fields competing for the same correspondence, whose differences should
be stated together.}
Figure~\ref{fig:hypergraph-set-system} illustrates this set-system view.

\noindent\textbf{Step 1 (Embeddings).}
For each column in a schema, we form a text representation from available metadata such as name and description, and compute an embedding using a pretrained encoder.

\noindent\textbf{Step 2 (Thresholded links).}
Given threshold \(\tau\) (equivalently, a distance cutoff \(\epsilon\)), we connect two columns \(x_i\) and \(x_j\) if \(\cos(e_i,e_j) \ge \tau\) (or distance \(\le \epsilon\)), producing a similarity link structure.
Building all thresholded links exactly requires comparing all pairs of columns, which is \(O(n^2)\) in the number of columns \(n\).
This full construction can be done offline; for larger schemas, approximate nearest-neighbor search (such as HNSW~\cite{malkov2018hnsw}) can be used to find neighbors above the threshold more efficiently (often roughly \(O(n\log n)\) in practice).

\noindent\textbf{Step 3 (Groups as confusable sets).}
\revadd{Several grouping rules can turn similarity signals into confusable
sets.  We use connected components under the thresholded links as a simple
default;} each component induces a
similarity group (a hyperedge/group) over columns.
In practice, the dominant cost is usually the LLM budget for producing group-wise differentiation cues; extracting groups from the link structure is lightweight by comparison.
\revadd{Section~\ref{sec:exp-results} evaluates alternative grouping rules.}

\subsection{\texorpdfstring{\mbox{Match-time}}{Match-time} materialization from a shortlist}
Although the hypergraph can be built offline, an exact full construction is \(O(n^2)\) and approximate indexes reduce the cost at the expense of approximation.
At match time (online), only a small portion of each schema is relevant to a given query, so we materialize it on demand.
\revadd{An embedding index retrieves relevant threshold neighborhoods without enumerating
every pair, and we form the induced groups for the current query.}

\begin{figure}[b]
\centering
\Description{A single-panel schematic of match-time use of similarity neighborhoods for target-side candidates. On the left, colored regions show threshold neighborhoods around candidates. On the right, the initial shortlist is C5 and C8; expansion adds C3, C4, C6, and C7, and connected components form groups for contrast cues.}
\includegraphics[width=\columnwidth]{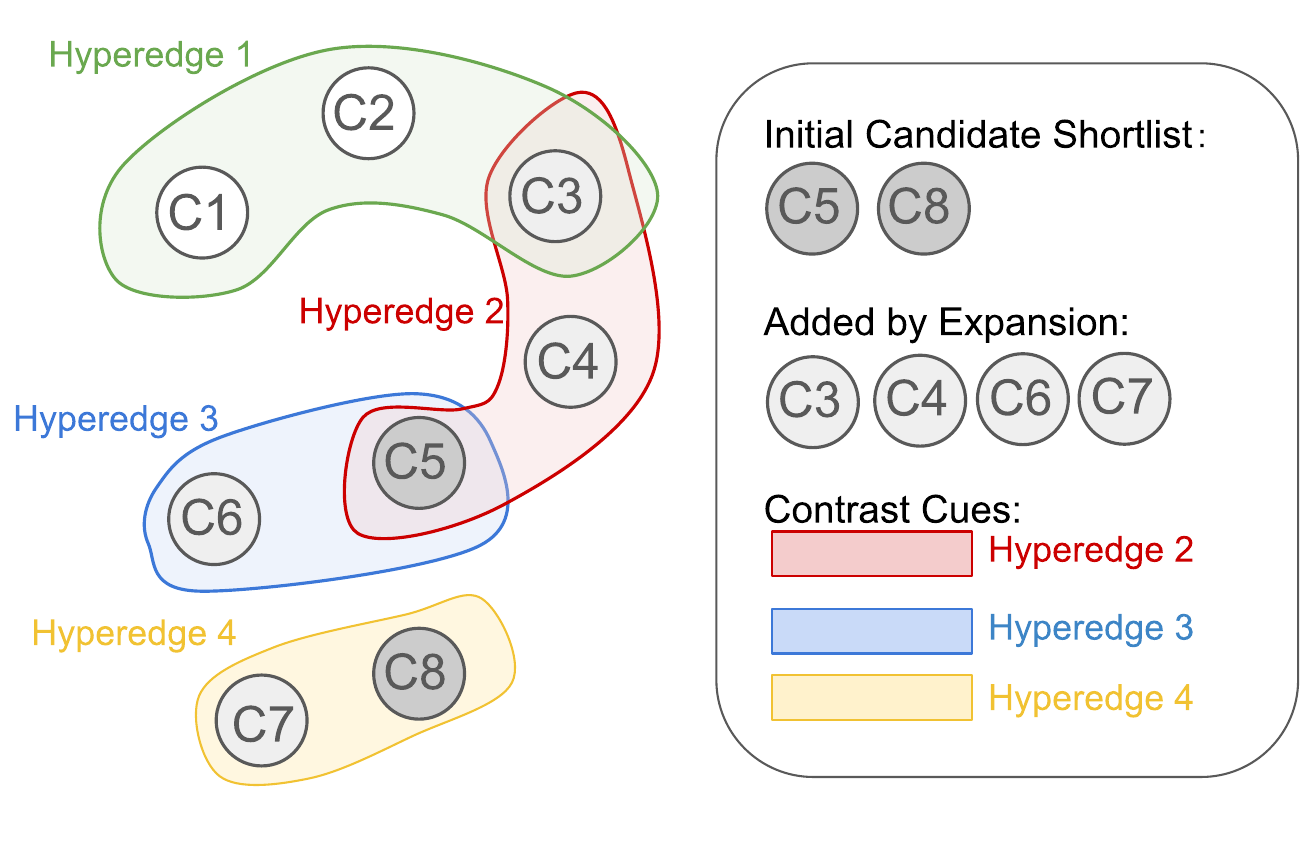}
\color{revisionred}
\revcaptionsetup
\caption{Match-time use of similarity groups from a target shortlist. The
initial shortlist \(\{C5,C8\}\) seeds neighborhoods that add
\(\{C3,C4,C6,C7\}\). Thresholded links in the resulting working set form
the connected-component groups used to generate contrast cues. The same
grouping-and-contrast procedure is applied to a capped, query-centered
neighborhood on the source side.}
\label{fig:similarity-hypergraph-overview}
\end{figure}

We use the similarity hypergraph in two \revadd{ways}.
\textbf{(i) Target-side candidate grouping.} Match-time processing starts from an upstream shortlist \(C_0\) (in our implementation, top-\(k\) by embedding cosine similarity to the source).
Expansion then scans \revadd{shortlist seeds, beginning with a small number of top-ranked candidates,} and adds columns that satisfy the same thresholded similarity rule used in the offline hypergraph construction.
These added columns are deduplicated against the original shortlist, and only a limited number are retained so that expansion does not pull in too many candidates.
The resulting working set \(C\) is then used to materialize thresholded links, whose connected components become the confusable groups for the query.
\textbf{(ii) Source-side query clarification.} We also materialize a small confusable set around the query column \(s\) in the \emph{source} schema \revadd{by pairing \(s\) with a capped number of its direct threshold neighbors} and generate a short differentiation block that highlights how \(s\) differs from its closest confusable source-side alternatives.
This is valuable when \(s\)'s description is generic and its intended meaning is only clear by contrasting it with nearby alternatives (Figure~\ref{fig:intro_time_example}).
For \revadd{the retained} non-singleton groups on either side, we generate group-wise contrast cues and append them to the final prompt so the LLM’s decision is made in direct comparison.
Figure~\ref{fig:similarity-hypergraph-overview} illustrates this match-time use of similarity groups.

\subsection{Grouped differentiation: generating contrastive cues}
Identifying a confusable group is only the first step; the final match still benefits from a brief, shared framing of \emph{how} its members differ.
ConStruM turns each \revadd{retained} non-singleton group (source-side confusable neighbors of the query, and target-side confusable groups within the candidate set) into a compact \emph{differentiation block} that can be dropped into the decision prompt.
To keep the contrasts grounded, the LLM sees each candidate’s metadata together with its tree-derived context, \shepadd{helping it explain differences in scope and meaning}.

The differentiation block follows a fixed format: a 1--2 sentence group summary (commonality + main axes of difference), followed by one short cue per candidate.
In practice, the block is generated on demand; \revadd{to bound the prompt, we cap the number of target groups and members that are differentiated}.

\section{Integration and Matching Pipeline}
\label{sec:pipeline}

\paragraph{\revadd{How ConStruM connects to a matcher}.}
ConStruM is not an end-to-end schema matcher.
It is an \emph{add-on} that strengthens the final, bounded-context decision in a local shortlist-and-decide pipeline.
At use time, we assume an upstream method proposes, for a query source column \(s\), a shortlist of target candidates \(C_0\subseteq T\).
The only requirement is that the upstream pipeline makes its final selection via one LLM call that chooses a single target from the shortlist.
ConStruM keeps the upstream candidate generation modular and instead supplies structured, query-specific evidence for that final decision.

\begin{revision}
\paragraph{Offline preprocessing (per dataset).}
Before queries arrive, ConStruM builds the reusable context tree
(Section~\ref{sec:tree}) and prepares the similarity-hypergraph layer
(Section~\ref{sec:group}).  It caches schema embeddings and builds an index
over target columns, supporting efficient match-time neighborhood lookup and
group materialization.
\end{revision}

\paragraph{Online matching (for one source column).}
\begin{description}
  \item[Upstream matcher (external).] Produce a shortlist \(C_0\subseteq T\) for \(s\).
  In our implementation, \(C_0\) is obtained by embedding retrieval (top-\(k\) targets by cosine similarity to \(s\)'s text metadata).
  More generally, \(C_0\) can come from ReMatch-style retrieval~\cite{sheetrit2024rematch} or from Matchmaker~\cite{seedat2024matchmaker}.

  \item[ConStruM add-on.] Given \(s\) and \(C_0\), ConStruM augments the final choice step:
  \begin{enumerate}
    \item \textbf{(Optional) Expand the working set.} Form a small candidate set \(C\) by adding a capped number of highly similar neighbors of strong candidates under a thresholded similarity rule (\(\tau\)).
    \item \textbf{Retrieve context packs.} Query the context tree to obtain a compact context pack for \(s\) and for each \(t\in C\) (column-to-root lineage summaries plus optional local relation snippets).
    \item \textbf{Add source-side contrast cues.} Materialize a small confusable group around \(s\) in the source schema and generate a short differentiation block (empty if no such group exists).
    \item \textbf{Add candidate-side contrast cues.} Use the target-side similarity hypergraph to materialize confusable groups within \(C\) and generate short differentiation blocks for \revadd{the retained} non-singleton groups (empty if no such groups exist).
    \item \textbf{Final LLM call.} Assemble one prompt containing the source metadata+context (plus any source differentiation block), each candidate’s metadata+context, and any candidate differentiation blocks; the LLM selects one target \(t\in C\).
  \end{enumerate}
\end{description}

\revadd{Appendix~\ref{app:prompt-formats} gives prompt templates and lower-level construction details.}

\section{Experiments}

\begin{revision}
We ask two questions.  First, can structured context resolve matches that
field metadata leaves ambiguous?  HRS-B and PSID-B provide the paper's main
accuracy comparison on difficult longitudinal-survey cases.  Second, can the
same add-on attach to a strong general matcher without disrupting it?
MIMIC--OMOP and Synthea--OMOP test this with a Matchmaker reconstruction.

Every method is asked to return one target column.  A decision is correct if that target
\shepadd{is an accepted match for the source field}.  HRS-B and PSID-B derive
\shepadd{these matches} from official cross-wave identifiers; the two general benchmarks use
fixed mappings.  These answers are visible only to the evaluator.

\subsection{Longitudinal context-stress benchmarks}
\label{sec:hrs}
\label{sec:psid}
We construct two context-stress benchmarks from the 
codebooks of
the Health and Retirement Study (HRS)~\cite{HRS} and Panel Study of Income
Dynamics (PSID)~\cite{psid}.  Each survey round, or \emph{wave}, has a
codebook whose organization reflects the questionnaire; we treat it as a schema.  We match fields from an earlier wave to
a later one.  Nearly identical field text can recur in
different sections or repeated questionnaire blocks.

These benchmarks grew out of a practical cross-wave harmonization task:
researchers need corresponding variables even as survey documentation
evolves.  Native HRS question identifiers are useful, but are not
consistently reliable across all sections and waves, which motivates a
semantic and contextual fallback. For unambiguous evaluation, HRS-B uses
Section J, where these identifiers are relatively reliable, while
PSID-B uses the official cross-year index.  Both label sources remain hidden
from every method.

HRS-B covers nine waves (2006--2022), 8,818 fields, and 26 earlier--later wave
pairs; PSID-B covers five waves (2007--2023), 2,742 fields, and all ten pairs.
One \emph{decision} matches one retained source field for one wave pair.
Because a source may be tested against several later waves, 92 distinct HRS
sources yield 190 decisions and 93 PSID sources yield 225.  These similar
decision counts conceal different construction workloads: ConStruM builds one
reusable tree over every field in each wave, not only the retained sources.

\paragraph{Benchmark construction.}
Selection concentrates on cases where a field's own metadata admits several
plausible targets; the official correspondence sources above determine the
answer only for evaluation.

Selection uses the earlier-wave codebook alone.  We find formulations with
whitespace-token Jaccard similarity at least 0.96 and exclude duplicated lines
from the same local block by requiring at least 16 fields of separation.  Both
members become source queries, with at most two from
any highly similar title group.  We retain a source only when its official
identifier supplies a later-wave match.  Target \shepadd{descriptions} and model outputs never
affect admission, and matching still searches the complete schemas.

The released schemas\footnote{\textcolor{revisionred}{HRS-B, PSID-B, and the
corresponding code are available at
\url{https://github.com/HoumingChen/construm}.}} retain the complete codebook organization but replace revealing variable
names with uniform wave-local IDs.  HRS preserves sanitized questionnaire
references and response counts; PSID masks native references and occurrence
markers, retains semantic value/range descriptions, and omits empirical
frequencies.  Otherwise they follow the same procedure.

Every scored source thus has a deliberately confusable counterpart in its
source schema, while the matcher still searches the complete later-wave
schema.  On PSID-B, 194/225
answers share their exact description with another target and 222/225 have a
target peer above 0.95 Jaccard similarity; a Jaccard-only top-1 matcher solves
56/225.

\subsection{Longitudinal evaluation protocol and methods}
\label{sec:baselines_expanded}
All 190 HRS-B and 225 PSID-B labels enter the denominator; retrieval misses,
invalid outputs, and failed calls count as incorrect.
\shepadd{On both surveys, ConStruM augments embedding retrieval followed by
LLM selection; the no-context condition evaluates this backbone without the
add-on.} The main comparison
uses \shepadd{eleven} method configurations:
\begin{itemize}
  \item \textbf{Embed-1NN}: select the embedding nearest neighbor (the
  highest-cosine target), without an LLM.
  \item \textbf{No-context LLM}: rerank the initial embedding shortlist using
  only source and candidate field metadata.
  \item \textbf{Tree only}: use ConStruM without its differentiation module;
  add tree-derived context packs and local windows.
  \item \textbf{Differentiation only}: use ConStruM without its tree module;
  add source- and target-side grouped differentiation and bounded candidate
  expansion.
  \item \textbf{ConStruM}: combine tree context and grouped differentiation.
\end{itemize}

\shepadd{Five additional} baselines provide broader context in different ways:
\begin{itemize}
  \item \textbf{ReMatch}: represent each field together with its
  ten-field neighborhood, retrieve the closest target neighborhood, and use
  the reconstructed ReMatch decision prompt to choose one of its fields
  ~\cite{sheetrit2024rematch}.
  \item \textbf{Dense+BM25 RAG} (retrieval-augmented generation): divide the
  target schema into eight-field chunks with four-field overlap, combine
  embedding and BM25 keyword ranks by reciprocal-rank
  fusion~\cite{robertson2009bm25}, and ask one LLM call to
  choose a field from the six highest-ranked chunks.
  \item \textbf{Column-graph RAG}: expand the 20 embedding candidates through
  target-column similarity links and nearby fields, then give the resulting
  subgraph to one LLM call over the original candidates.  This is a
  column-graph retrieval baseline, not GraphRAG's extracted knowledge graph
  and community-summary pipeline.
  \item \textbf{Full-schema LLM}: give one forced-choice call all field
  identifiers and descriptions from both schemas.  The model may return any
  target-schema field; no shortlist or retrieved evidence is supplied.
  \item \textbf{LC-1/3} (long-context thirds): split each schema into three
  consecutive parts.  Pair the query's source part with each target part in
  three proposal calls, then choose among their proposals in a fourth call.
  The full target schema is therefore exposed in pieces.
\end{itemize}
\begin{shepherding}
The full \textbf{Matchmaker reconstruction}
~\cite{seedat2024matchmaker}, including ColBERT and LLM-based candidate
retrieval, refinement, confidence scoring, and self-improvement with synthetic
demonstrations, tests whether these stages can resolve the same contextual
ambiguities. Because these benchmarks mask native column names, we generate
short names from each field's description alone, without surrounding fields
or correspondence labels. Identical descriptions may receive identical names;
selecting a name retrieves all fields with that name before refinement.
The refined shortlist contains 20 candidates, matching our initial retrieval
budget, and all LLM stages use GPT-5.4, as in the other LLM-based survey methods.
\end{shepherding}

All \shepadd{eleven} methods cover every decision in both benchmarks.  Here and below, a
\emph{paired comparison} evaluates two methods on the same decisions; its wins
and losses count decisions correct only by the first method versus only by the
second.
Appendix~\ref{app:parameters} gives the exact graph and partition settings.

\paragraph{Models and key hyperparameters.}
Longitudinal experiments use \textbf{GPT-5.4}
(\nolinkurl{gpt-5.4-2026-03-05}) and
\nolinkurl{text-embedding-3-small}; \shepadd{HRS-B} ReMatch retains its
\nolinkurl{text-embedding-ada-002} retriever.  We retrieve \(k=20\) candidates
and allow bounded expansion to add at most 12 similarity neighbors, for a
final cap of 32.  This leaves room for confusable alternatives while bounding
the decision prompt.  We group confusable fields at cosine similarity
\(\tau=0.95\).  One context tree per wave is reused across its decisions.
Appendix~\ref{app:parameters} records the remaining user-facing settings.

\begin{table*}[t]
  \centering
  \color{shepherdred}
  \captionsetup{labelfont={color=shepherdred},textfont={color=shepherdred}}
  \caption{Mean forced-choice accuracy (\%) [pointwise 95\% confidence
  interval] over three runs. \emph{Combined} counts all 415 decisions from
  the two benchmarks equally. For combined accuracy, $p_{\mathrm H}$ is the
  Holm-adjusted $p$-value for each method's two-sided Welch--Satterthwaite
  $t$-test against ConStruM (ten comparisons).}
  \label{tab:longitudinal-results}
  \small
  \setlength{\tabcolsep}{5pt}
  \begin{tabular}{lcccc}
    \toprule
    & \textbf{HRS-B (190)} & \textbf{PSID-B (225)}
      & \multicolumn{2}{c}{\textbf{Combined (415)}} \\
    \cmidrule(lr){4-5}
    \textbf{Method} & \textbf{Accuracy} & \textbf{Accuracy}
      & \textbf{Accuracy} & $\boldsymbol{p_{\mathrm H}}$ \\
    \midrule
Embed-1NN & 33.2 [33.2, 33.2] & 33.8 [33.8, 33.8] & 33.5 [33.5, 33.5] & $1.77\times10^{-4}$ \\
No-context LLM & 54.0 [52.0, 56.0] & 49.6 [49.0, 50.3] & 51.6 [50.8, 52.4] & $3.13\times10^{-6}$ \\
ReMatch reimplementation & 39.5 [38.2, 40.8] & 33.8 [33.8, 33.8] & 36.4 [35.8, 37.0] & $2.43\times10^{-5}$ \\
Matchmaker reconstruction & 60.5 [58.3, 62.8] & 29.9 [29.3, 30.6] & 43.9 [43.0, 44.9] & $9.06\times10^{-7}$ \\
Dense+BM25 RAG & 44.0 [43.3, 44.8] & 36.9 [35.0, 38.8] & 40.2 [39.2, 41.1] & $7.86\times10^{-7}$ \\
Column-graph RAG & 61.8 [59.8, 63.8] & 39.0 [37.3, 40.6] & 49.4 [48.6, 50.2] & $1.04\times10^{-7}$ \\
Full-schema LLM & 96.1 [90.7, 100.0] & 74.2 [68.4, 80.1] & 84.3 [81.6, 86.9] & $3.83\times10^{-4}$ \\
LC-1/3 & 92.6 [86.6, 98.6] & 62.1 [61.4, 62.7] & 76.1 [73.4, 78.7] & $2.58\times10^{-4}$ \\
Tree only & 97.4 [97.4, 97.4] & 68.4 [63.6, 73.3] & 81.7 [79.1, 84.3] & $3.83\times10^{-4}$ \\
Differentiation only & 97.5 [95.5, 99.5] & 81.3 [76.9, 85.7] & 88.8 [86.7, 90.8] & $5.78\times10^{-4}$ \\
\textbf{ConStruM} & \textbf{100.0 [100.0, 100.0]} & \textbf{91.6 [88.6, 94.5]} & \textbf{95.4 [93.8, 97.0]} & --- \\
    \bottomrule
  \end{tabular}
\end{table*}

\paragraph{Evaluation.}
\begin{shepherding}
To assess sensitivity to LLM variability, we evaluate each of the eleven methods
three times. Each repetition uses fresh matching-stage LLM responses,
including differentiation cues where applicable, while reusing trees and
embeddings. Matchmaker also regenerates its synthetic demonstrations,
candidate reasoning, and refinement, while reusing description-derived names
and semantic retrieval.

Table~\ref{tab:longitudinal-results} reports mean accuracy and approximate
95\% confidence intervals. Combined accuracy counts all 415 decisions from
the two benchmarks equally.

For combined accuracy, we use two-sided unequal-variance $t$-tests
(Welch--Satterthwaite)~\cite{welch1947generalization,satterthwaite1946approximate}.
The tests and confidence intervals use a separate variance estimate for each
method and benchmark. Holm correction controls false positives across
the ten combined ConStruM-versus-alternative comparisons~\cite{holm1979simple}.
These tests concern overall accuracy on the merged benchmark, not superiority
on each survey separately. They assume independent matching repetitions and
approximately normal run scores; three runs limit precision. The intervals describe variability on
these fixed benchmarks with the stated reusable preprocessing. Equal scores
across three runs indicate no observed variation, not guaranteed determinism.
Appendix~\ref{app:repeat-statistics} provides the full calculation and secondary benchmark-specific tests.

The smaller-model comparison follows the same repetition procedure.
The supporting grouping, construction, and fixed-candidate
comparisons report single-run results with exact paired McNemar tests.
Single-run results by wave pair are included in Appendix~\ref{app:hrs_b_pairwise}.
\end{shepherding}
Table~\ref{tab:longitudinal-results} evaluates each method as described above;
Section~\ref{sec:matched-controls} later fixes the candidate list and decision
prompt to compare the evidence itself.

\paragraph{Cost measurement.}
We meter provider-reported uncached input, cached input, and output tokens for
every attempt.  On 2026-08-09, GPT-5.4 charged \$2.50, \$0.25, and \$15 per
million tokens, respectively; above 272K input tokens, input cost doubles and
output cost rises by 1.5\(\times\).  Embedding inputs are charged at \$0.02 per
million tokens for \nolinkurl{text-embedding-3-small} and \$0.10 per million
for \nolinkurl{text-embedding-ada-002}.  We separate reusable construction from online
matching, charging each tree once before amortizing it over its decisions.

\begin{table*}[t]
  \centering
  \color{revisionred}
  \caption{Measured API use for \shepadd{single evaluations of ten methods
  from} Table~\ref{tab:longitudinal-results}.
  Input is online chat input (cached plus uncached), in millions of tokens.
  Calls include preparation and online requests; dollar costs separate reusable
  preparation from online matching.  One embedding request may batch many
  fields; methods are run separately.  Components are rounded separately;
  totals use unrounded costs.}
  \label{tab:cost-comparison}
  \footnotesize
  \renewcommand{\arraystretch}{.92}
  \setlength{\tabcolsep}{2.4pt}
  \begin{tabular*}{\textwidth}{@{\extracolsep{\fill}}lrrrrrrrrrr@{}}
    \toprule
      & \multicolumn{5}{c}{\textbf{HRS-B (190 decisions)}}
      & \multicolumn{5}{c}{\textbf{PSID-B (225 decisions)}} \\
    \cmidrule(lr){2-6}\cmidrule(lr){7-11}
    \textbf{Method} & \textbf{Calls} & \textbf{Input} & \textbf{Preparation}
      & \textbf{Online} & \textbf{Total}
      & \textbf{Calls} & \textbf{Input} & \textbf{Preparation}
      & \textbf{Online} & \textbf{Total} \\
      & & \textbf{(M)} & \multicolumn{3}{c}{\textbf{Cost (\$)}}
      & & \textbf{(M)} & \multicolumn{3}{c}{\textbf{Cost (\$)}} \\
    \midrule
    Embed-1NN & 73 & 0 & 0.02 & 0 & 0.02 & 25 & 0 & 0.01 & 0 & 0.01 \\
    No-context LLM & 263 & 0.59 & 0.02 & 1.85 & 1.87 & 251 & 0.81 & 0.01 & 2.41 & 2.42 \\
    ReMatch reimplementation & 219 & 0.60 & 0.24 & 1.46 & 1.69 & 235 & 1.17 & 0.03 & 2.79 & 2.82 \\
    Dense+BM25 RAG & 252 & 1.45 & 0.09 & 3.68 & 3.76 & 245 & 2.91 & 0.04 & 7.51 & 7.55 \\
    Column-graph RAG & 265 & 2.91 & 0.02 & 7.32 & 7.33 & 250 & 4.82 & 0.01 & 12.29 & 12.30 \\
    Full-schema LLM & 190 & 51.12 & 0 & 30.90 & 30.90 & 225 & 44.57 & 0 & 15.12 & 15.12 \\
    \textsc{LC-1/3} & 760 & 50.50 & 0 & 74.86 & 74.86 & 900 & 46.57 & 0 & 77.19 & 77.19 \\
    Tree only & 1328 & 6.70 & 25.86 & 16.76 & 42.62 & 550 & 11.04 & 9.76 & 20.11 & 29.87 \\
    Differentiation only & 447 & 1.44 & 0.02 & 4.73 & 4.74 & 424 & 2.48 & 0.01 & 8.00 & 8.01 \\
    ConStruM & 1508 & 10.50 & 25.86 & 12.91 & 38.77 & 796 & 15.43 & 9.76 & 37.98 & 47.74 \\
    \bottomrule
  \end{tabular*}
\end{table*}
\end{revision}

\begin{revision}
\subsection{Longitudinal results and analysis}
\label{sec:exp-results}
\subsubsection{Longitudinal context-stress results}
\label{sec:psid-results}
\begin{shepherding}
ConStruM ranks first on both benchmarks in all three evaluations
(Table~\ref{tab:longitudinal-results}). The large gains over field-only matching
persist across runs. On the merged 415-decision benchmark, its mean accuracy is
significantly higher than every alternative after correction (largest adjusted
$p=5.78\times10^{-4}$), including Matchmaker, full-schema prompting, and LC-1/3. This supports the overall
advantage of selecting and organizing schema evidence across the two surveys.
\end{shepherding}

With field metadata alone, roughly half the decisions remain incorrect.
Broader-context retrieval baselines recover only part of this gap.  Both
structured modules outperform no context, and their combination ranks first.

\subsubsection{Cost}
Table~\ref{tab:cost-comparison} \shepadd{separates reusable preparation from
online matching costs.}

\shepadd{Across both benchmarks, full-schema} prompting transmits 3.69 times as many online input tokens as
ConStruM, and 92 HRS-B prompts exceed 272K tokens.  \shepadd{Prefix caching
reduces full-schema prompting's bill without shortening its prompts.
ConStruM makes more calls and has a higher total API cost on both benchmarks,
trading that additional expense for the accuracy advantage in
Table~\ref{tab:longitudinal-results}.}

\end{revision}

\begin{revision}
\subsubsection{Grouping choices and smaller-model evaluation}

We change only the target-side grouping rule, keeping the decisions, candidate
lists, source groups, trees, model, and prompt templates fixed.  Because full HRS-B is
at its ceiling, we also test differentiation without tree context, where the
effect of grouping is easier to see.  Table~\ref{tab:grouping-robustness}
compares nine rules, including nearest-neighbor
components, complete and average linkage~\cite{murtagh2012algorithms},
DBSCAN~\cite{ester1996dbscan}, overlapping neighborhoods, and threshold
components; distance .05 equals similarity .95.

With any of the eight alternative grouping rules, full accuracy stays at
189--190/190 on HRS-B and 203--206/225 on PSID-B; none differs from the $.95$
default after Holm correction for the multiple tests.  Accuracy is therefore
insensitive to the particular grouping algorithm among those tested.
At $.93$, HRS
differentiation-only is one decision higher than at $.95$; both full
evaluations tie.  The one-decision gain is not statistically significant, so
it provides no evidence that $.93$ is better.  Thus the results are not tied
to one reasonable grouping rule.  We retain $.95$ threshold components:
they were the simple default before this sensitivity comparison and remain
consistent across all three panels.

\begin{table}[t]
  \centering
  \color{revisionred}
  \caption{Target-side grouping sensitivity.  \emph{HRS diff.} uses
  differentiation without tree context.}
  \label{tab:grouping-robustness}
  \footnotesize
  \setlength{\tabcolsep}{1.8pt}
  \begin{tabular}{@{}p{.34\columnwidth}rrr@{}}
    \toprule
    \textbf{Grouping rule} & \textbf{HRS full (190)} & \textbf{HRS diff. (190)} &
      \textbf{PSID full (225)} \\
    \midrule
    Mutual 3-NN components & 189/190 (99.5\%) & 174/190 (91.6\%) & 204/225 (90.7\%) \\
    Complete link ($.05$) & 190/190 (100.0\%) & 178/190 (93.7\%) & 204/225 (90.7\%) \\
    Average link ($.05$) & 190/190 (100.0\%) & 178/190 (93.7\%) & 205/225 (91.1\%) \\
    DBSCAN ($.05$, min 3) & 190/190 (100.0\%) & 184/190 (96.8\%) & 206/225 (91.6\%) \\
    Overlap neighborhoods ($.95$) & 190/190 (100.0\%) & 185/190 (97.4\%) & 203/225 (90.2\%) \\
    Threshold components, $.90$ & 189/190 (99.5\%) & 184/190 (96.8\%) & 205/225 (91.1\%) \\
    Threshold components, $.93$ & 190/190 (100.0\%) & 186/190 (97.9\%) & 205/225 (91.1\%) \\
    Threshold components, $.95$ (default) & 190/190 (100.0\%) &
      185/190 (97.4\%) & 205/225 (91.1\%) \\
    Threshold components, $.97$ & 190/190 (100.0\%) & 181/190 (95.3\%) & 205/225 (91.1\%) \\
    \bottomrule
  \end{tabular}
\end{table}

Table~\ref{tab:smaller-model} tests whether ConStruM remains useful with a
smaller model.  GPT-5.4-mini is evaluated end to end under the no-context and
full-ConStruM conditions.  No context chooses from the retrieved top 20;
ConStruM uses its bounded candidate expansion and structured evidence, with
trees and differentiation cues built by GPT-5.4-mini itself.
Both mini arms use a compact decision template and allow one label-blind
retry for an out-of-roster answer, with unchanged candidates and evidence;
any remaining invalid answer counts as incorrect.

\begin{table}[htbp]
  \centering
  \color{revisionred}
  \caption{Effect of ConStruM with GPT-5.4 and GPT-5.4-mini on HRS-B.
  \shepadd{Mean accuracy (\%) [95\% confidence interval] over three runs.}}
  \label{tab:smaller-model}
  \small
  \setlength{\tabcolsep}{6pt}
  \begin{tabular}{lcc}
    \toprule
    \textbf{Model} & \textbf{No context} & \textbf{ConStruM} \\
    \midrule
    GPT-5.4 & \shepadd{54.0 [52.0, 56.0]} & \shepadd{100.0 [100.0, 100.0]} \\
    GPT-5.4-mini & \shepadd{40.9 [39.4, 42.4]} & \shepadd{48.8 [42.3, 55.2]} \\
    \bottomrule
  \end{tabular}
\end{table}

\shepadd{ConStruM increases mini's mean accuracy by 7.9 percentage points
(two-sided Welch $t$-test, $p=.0289$), compared with 46.0 points for GPT-5.4.}
Structured context therefore helps both models,
although GPT-5.4-mini's absolute accuracy remains lower.

\subsubsection{Context-tree construction ablation}

Table~\ref{tab:construction-ablation} ablates tree construction on PSID-B's
largest wave pair, 2019--2023 (1,189 fields; 27 decisions).  Only tree construction changes: candidate
lists, the decision template, and non-tree inputs stay fixed, and differentiation is disabled.
Stages 1--4 summarize
local windows, infer theme and flow, synthesize a conceptual map, and refine
region boundaries (Section~\shepadd{\ref{sec:within-table-construction}}).  Fixed width instead
nests equal-sized regions, while Stage 1 only stops after local grouping.
Stage 2 is omitted by supplying no global summary to the synthesis stage.
Without Stage 4, a fixed midpoint rule
turns Stage-3 rough intervals into contiguous regions.  We do not remove Stage
3 in isolation because Stage 4 consumes its synthesized map; doing so would
require inventing a replacement procedure rather than ablating the method.

\begin{table}[htbp]
  \centering
  \color{revisionred}
  \caption{Context-tree construction ablation on PSID-B 2019--2023 (27
  decisions), with differentiation disabled.  Costs cover tree construction
  and all decisions.  Components are rounded
  separately; totals use unrounded costs.}
  \label{tab:construction-ablation}
  \small
  \setlength{\tabcolsep}{3pt}
  \begin{tabular}{lrrrr}
    \toprule
    \textbf{Construction} & \textbf{Correct} & \textbf{Build} &
      \textbf{Online} & \textbf{Total} \\
    \midrule
    No tree & 16/27 (59.3\%) & \$0.00 & \$1.66 & \$1.66 \\
    Fixed width & 19/27 (70.4\%) & \$0.99 & \$2.53 & \$3.53 \\
    Stage 1 only & 16/27 (59.3\%) & \$1.87 & \$2.14 & \$4.00 \\
    Full minus Stage 2 & 24/27 (88.9\%) & \$3.62 & \$2.45 & \$6.08 \\
    Full minus Stage 4 & 20/27 (74.1\%) & \$2.91 & \$2.46 & \$5.37 \\
    \textbf{Full Stages 1--4} & \textbf{27/27 (100.0\%)} & \$3.53 & \$2.65 & \$6.18 \\
    \bottomrule
  \end{tabular}
\end{table}

These results support the complete four-stage construction as the general
default.  It is the only design to solve all 27 decisions, and every coherent
reduction except omitting Stage 2 is significantly worse after Holm correction
(adjusted \(p\leq.0313\); Stage-2 omission \(p=.25\)).  A simpler construction
may suffice for some schemas, but retaining all four stages gives the strongest
result without requiring a schema-specific choice among reduced designs.

\subsubsection{Isolating evidence selection}
\label{sec:matched-controls}

Table~\ref{tab:longitudinal-results} answers the end-to-end question: can a
recent long-context model replace ConStruM by reading the complete schemas?
ConStruM is more accurate on both benchmarks.  Because the two complete
pipelines also form their decisions differently, we add a controlled
comparison to isolate the value of evidence selection itself.

Both conditions in Table~\ref{tab:selected-context-controls} reuse the same
target roster, \shepadd{capped at 32 candidates}, produced by the common configuration in
Section~\ref{sec:baselines_expanded}.  They also use the same model, decision
instruction, and output rule.  One adds ConStruM's tree and differentiation
evidence; the other instead adds every source and target field.  Thus only the
\shepadd{schema evidence supplied for the decision} changes.  Because this
control \shepadd{restricts the candidate list}, its accuracies need not equal the
unrestricted full-schema row in Table~1.

\begin{table}[!htbp]
  \centering
  \color{revisionred}
  \caption{Selected versus all-fields evidence \shepadd{with the same candidate
  list, capped at 32}.  ``Wins--losses'' reports selected-only versus all-fields-only
  successes; $p$ is from the exact McNemar test.
  \emph{Input} is online input tokens; \emph{Long} counts calls above 272K
  input tokens; and \emph{Bill} is the measured charge after caching.}
  \label{tab:selected-context-controls}
  \footnotesize
  \setlength{\tabcolsep}{2pt}
  \begin{tabular}{@{}lrrrr@{}}
    \toprule
    \textbf{Benchmark} & \textbf{Selected} & \textbf{All fields} & \textbf{Wins--losses} & \textbf{$p$} \\
    \midrule
    HRS-B & 186/190 (97.9\%) & 158/190 (83.2\%) & 30--2 & $2.46\!\times\!10^{-7}$ \\
    PSID-B & 213/225 (94.7\%) & 188/225 (83.6\%) & 27--2 & $1.62\!\times\!10^{-6}$ \\
    Combined & 399/415 (96.1\%) & 346/415 (83.4\%) & 57--4 & $4.85\!\times\!10^{-13}$ \\
    \addlinespace
    \multicolumn{5}{@{}l}{\emph{Resource use across both benchmarks}} \\
    \cmidrule(l){1-5}
    \textbf{Evidence} & \textbf{Input} & \textbf{Long} & \multicolumn{2}{c}{\textbf{Bill}} \\
    Selected context & 19.68M & 0 & \multicolumn{2}{c}{\$45.56} \\
    All fields & 98.73M & 112 & \multicolumn{2}{c}{\$59.29} \\
    \bottomrule
  \end{tabular}
\end{table}

\noindent\textbf{Accuracy.}
\shepadd{Selected context is more accurate on both benchmarks under the common
decision interface.} The
advantage therefore comes from selecting and organizing useful evidence, not
from different candidate rosters or decision wording.

\noindent\textbf{Prompt size and model availability.}
The all-fields condition transmits 5.02 times as many input tokens, and 112
HRS-B calls exceed 272K tokens; no selected-context call does.

\end{revision}

\begin{revision}
\subsection{General schema-matching benchmarks and results}
\label{sec:mimic}
MIMIC--OMOP and Synthea--OMOP map two source schemas to the Observational
Medical Outcomes Partnership (OMOP) common data model
~\cite{overhage2012omopcdm}.  The first maps the MIMIC-III schema introduced in
Figure~\ref{fig:intro_time_example}~\cite{sheetrit2024rematch}; the second maps
a schema generated by Synthea to the same target.  Under the forced-choice
evaluation used here, they contain 155 and 38 decisions, respectively.  A
Synthea source may accept several targets, but each prediction still names one.

Matchmaker reports $62.2{\pm}2.4\%$ top-1 accuracy on MIMIC--OMOP
~\cite{seedat2024matchmaker}.  Its executable pipeline is not publicly
released, and the paper leaves several operational choices unspecified, so we
could not exactly reproduce that result from the available artifacts.  We
nevertheless implement every described stage---candidate retrieval and
refinement, confidence scoring, and self-improvement with synthetic
demonstrations---and make consistent, best-effort choices where details are
unavailable.  Following the paper's use of GPT-4-family models, we use
\textbf{GPT-4.1} throughout our reconstruction.  It reaches 61.3\%, close to the
published result.  Base and add-on arms share its candidates and scores;
ConStruM adds evidence only to the final choice, while everything upstream
remains unchanged.

\subsubsection{Results}
Table~\ref{tab:mimic_top1} reports the primary add-on comparison on both
general schema-matching benchmarks.

\begin{table}[!htbp]
  \centering
  \color{revisionred}
  \caption{ConStruM as a final-stage add-on to the Matchmaker reconstruction
  on general schema-matching benchmarks.  Upstream candidates and
  confidence scores are fixed within each comparison.  Published Matchmaker
  accuracy is included for reference; the exact McNemar test compares the two
  reconstruction columns.}
  \label{tab:mimic_top1}
  \small
  \setlength{\tabcolsep}{2.2pt}
  \begin{tabular}{lrrrr}
    \toprule
    \textbf{Benchmark} & \textbf{Published} & \textbf{Reconst.} &
      \textbf{+ConStruM} & \textbf{\(p\)} \\
    \midrule
    MIMIC--OMOP & $62.2{\pm}2.4\%$ & 95/155 (61.3\%) & 96/155 (61.9\%) & 1.000 \\
    Synthea--OMOP & $70.2{\pm}1.7\%$ & 26/38 (68.4\%) & 26/38 (68.4\%) & 1.000 \\
    \bottomrule
  \end{tabular}
\end{table}

Adding ConStruM changes MIMIC--OMOP accuracy from 61.3\% to 61.9\%; on
Synthea--OMOP, both arms reach 68.4\%.  Neither paired difference is
statistically significant.  \shepadd{One possible explanation is that}
Matchmaker's field and table metadata, candidate set, and confidence scores
\shepadd{leave little ambiguity that additional context can resolve.}  It may help a few
remaining ambiguities while adding small distractions elsewhere; these samples
cannot separate such small opposing effects, but the measured net change is
much smaller than on HRS-B and PSID-B.
\end{revision}

\noindent

\begin{revision}
\section{Discussion}
\label{sec:discussion}
\paragraph{Where structured context helps.}
HRS-B and PSID-B concentrate on cases where field metadata leaves several
plausible correspondences.  The tree supplies schema scope, differentiation
contrasts the remaining alternatives, and their combination ranks first on
both independently maintained surveys.  The full-schema comparisons sharpen
this result: even when a recent long-context model can accept every field,
presenting everything is less accurate and substantially less compact than
selecting and organizing the evidence.  MIMIC--OMOP and Synthea--OMOP mark the
other end of the spectrum.
On those two general benchmarks, adding ConStruM produces no significant
accuracy change.  The combined results therefore place its measured benefit
in context-sensitive matching, while showing little effect in the general
settings tested.
Separately, the grouping sensitivity study shows that several reasonable
grouping rules yield similar full-pipeline accuracy.
\shepadd{This supports the use of a simple grouping rule to identify which
alternatives should be contrasted.}

\paragraph{Practical use.}
The schema structures are built once and reused, so deployment is most
attractive for large, repeatedly matched, or evolving schemas.  It can also be
invoked selectively when a shortlist contains near-duplicates or has a small
confidence margin, retaining the original path for easy cases.  This paper
evaluates single-target forced choice; calibrated rejection and complex
one-to-many mappings remain important extensions.  A further direction is to
detect when additional context is needed and adapt the evidence budget to the
difficulty of each decision.
\end{revision}

\section{Related Work}

\noindent\textbf{Classical and embedding-based schema matching.}
\shepadd{Classical schema matchers}~\cite{rahm2001survey}, \shepadd{such as}
COMA/COMA++~\cite{Do2002COMAA}, \shepadd{combine linguistic, constraint-based,
and structural signals. Related work addresses uncertainty and aggregation of
match evidence}~\cite{Gal2011UncertainSM}.
More recently, embedding-based similarity and neural encoders have been used for scalable matching and candidate generation~\cite{zhang2021smat,Shraga2020ADnEVCS,Li2020DeepEM}.

\noindent\textbf{Task-specific neural models (no external LLM).}
Between classical matchers and prompt-based LLM systems, several lines of work use deep learning without relying on an \emph{external} general-purpose LLM at inference time.
This includes task-specific deep architectures for schema matching and related integration problems, including graph/relational embedding approaches such as REMA~\cite{koutras2020rema}.
Learned Schema Mapper (LSM) \shepadd{uses} pre-trained language models as linguistic encoders together with active learning and informative attribute selection to reduce labeling cost in enterprise schema mapping~\cite{zhang2023plm}.
Unicorn~\cite{tu2023unicorn} \shepadd{learns} a unified supervised model across multiple matching tasks, including schema matching, via a shared encoder and mixture-of-experts architecture rather than prompt-based decision making.
Jellyfish~\cite{zhang-etal-2024-jellyfish} \shepadd{provides} a \emph{locally deployable} model instruction-tuned for data preprocessing tasks, including schema matching, rather than prompting a closed, off-the-shelf LLM service.

\noindent\textbf{Prompted LLM pipelines (retrieval-augmented and multi-stage).}
Recent work explores how to prompt LLMs for schema matching, and shows that both too little and too much context can hurt performance~\cite{parciak2024schema}.
ReMatch~\cite{sheetrit2024rematch} uses retrieval to select target tables before an LLM makes column-level decisions, improving scalability compared to exhaustive pairwise prompting.
Matchmaker~\cite{seedat2024matchmaker} proposes a compositional, multi-stage LLM program that self-improves via synthetic demonstrations.
Schemora~\cite{gungor2025schemora} emphasizes multi-stage recommendation and metadata enrichment for LLM-based schema matching under practical retrieval constraints.
Other contemporary systems study complementary directions, including knowledge-compliant reasoning~\cite{xu2024kcmf} and security-/privacy-aware retrieval augmentation~\cite{liu2024gram}, as well as modular multi-stage pipelines~\cite{wang2025llmatch} and hybrid feature + generation approaches~\cite{zhang2024smutf}.
Magneto~\cite{liu2025magneto} combines smaller models for efficient candidate selection with larger models for reranking.
ConStruM \shepadd{provides} a \emph{structure-guided context module} that can be attached to an existing candidate-selection or reranking step to improve the \emph{decision prompt} under a fixed budget.

\begin{revision}
\noindent\textbf{Mechanisms for difficult pairs.}
The mechanism for resolving a difficult pair has changed with the matching
tools available.  Classical schema matchers combine several schema signals;
Similarity Flooding, for example, propagates consistency through neighboring
relations~\cite{melnik2002similarity}.  Supervised neural entity matchers learn
pair representations from labeled examples: Ditto augments difficult
examples, while Battleship spends a limited labeling budget on informative
tuple pairs and retrains iteratively
~\cite{Li2020DeepEM,genossar2023battleship}.  Contemporary LLM schema matchers
change the intervention point again: ReMatch and Matchmaker narrow or refine
the candidates and use LLM reasoning to make the final choice, while Magneto
combines small-model retrieval with a larger-model reranker
~\cite{sheetrit2024rematch,seedat2024matchmaker,liu2025magneto}.

Within this contemporary pipeline, ConStruM acts as an add-on to the LLM
decision.  It is particularly useful when the correct target and several
alternatives look highly similar from their names and descriptions.  Its global
similarity hypergraph groups highly similar attributes in the source and
target schemas; on the target side, it can also add a similar attribute that
the initial candidate list omitted.  ConStruM then produces group-wise cues
that state how these attributes differ, while the context tree supplies the
surrounding schema information needed to interpret those differences.  The
existing LLM selector makes the final choice from this enriched prompt.
Section~\ref{sec:mimic} tests this relationship directly by
attaching ConStruM to our Matchmaker reconstruction.
\end{revision}

\noindent\textbf{KG-augmented LLM pipelines.}
Some LLM-based matchers augment prompts with \emph{external} structured resources, especially knowledge graphs.
For example, KG-RAG4SM \cite{ma2025knowledge} studies knowledge-graph-based retrieval-augmented generation for schema matching, and SMoG \cite{jeon2025smog} explores iterative graph exploration for efficient and explainable integration.
These directions primarily focus on how to \emph{use existing external graphs} to ground or explain matching decisions.
In contrast, ConStruM mines the \emph{internal} structure of the schema itself (via reusable schema-side hierarchies and similarity graphs) to build compact, query-specific context packs.

\noindent\textbf{Hierarchical and graph-based retrieval for structured context.}
Graph-based RAG systems such as GraphRAG~\cite{edge2024graphrag} and LightRAG~\cite{guo2025lightrag} highlight the value of organizing evidence as a graph and retrieving a query-specific subgraph for holistic, multi-hop context.
Complementarily, hierarchical retrieval methods such as RAPTOR~\cite{sarthi2024raptor} represent documents at multiple granularities via tree-structured summaries.
\begin{revision}
The shared inspiration is the broad systems principle of organizing a large
evidence space before retrieval; the connection is conceptual rather than
algorithmic.
RAPTOR recursively clusters and summarizes document chunks for question
answering, while GraphRAG extracts an entity graph and hierarchical community
summaries for corpus-level sensemaking.  ConStruM instead begins with two
schemas, one source field, and an upstream target shortlist.  That decision
structure determines both of its components: the context tree preserves field
identity and native schema scope, including a four-stage construction for wide
tables, while the similarity hypergraph forms sets of fields that compete for
the same correspondence and generates group-wise contrasts.  These are
schema-matching operations, not document clustering or entity-community
summarization, and their output is evidence for one forced-choice match rather
than an answer to an open-ended document query.  The simpler idea of retrieving
a query-centered column graph is evaluated separately as column-graph RAG; it
does not include ConStruM's schema-specific hierarchy or contrast generation.
\end{revision}

\noindent\textbf{Different notions of context in schema matching.}
Prior work studies \emph{contextual} matching where a correspondence is paired with \emph{selection conditions} (predicates) under which it is semantically valid, for example in Bohannon et al.~\cite{bohannon2006context}.
\revadd{There, context is part of the mapping itself: the predicates specify the
circumstances in which a stated correspondence applies.}
\revadd{We retain the term because surrounding schema information plays the
familiar role of context: it helps interpret an otherwise ambiguous field.}
ConStruM uses \emph{context} to mean a budgeted, query-specific evidence pack (our \emph{context pack}; multi-level neighborhood/scope summaries, lightweight relations, and contrast cues) that supports an LLM’s disambiguation among confusable candidates.
\revadd{The surrounding schema evidence helps identify a fixed correspondence;
it does not define conditions under which that correspondence changes.}

\noindent\textbf{Other context-heavy data tasks.}
Table-understanding models such as TURL focus on \emph{within-table} tasks including entity linking, column typing, and cell filling~\cite{deng2021turl}, while Starmie studies data-lake discovery via contextualized column embeddings for \emph{table union search}~\cite{fan2023starmie}.
These directions show that context matters, but they target table interpretation or retrieve-and-score unionability over value-populated tables rather than \emph{cross-schema} correspondence decisions over schema metadata under an LLM \emph{shortlist} and \emph{fixed prompt budget}.
ConStruM instead focuses on selecting and organizing query-specific evidence so that a final LLM call can disambiguate semantically close target attributes.

\begin{shepherding}
\section{Limitations}
\label{sec:limitations}
ConStruM improves both tested LLMs (Table~\ref{tab:smaller-model}). However,
on HRS-B, GPT-5.4-mini with ConStruM still falls below GPT-5.4 without
context. Its gains may therefore be insufficient to overcome a large
difference in model capability. ConStruM remains useful for improving a
given model: with GPT-5.4, it raises mean HRS-B accuracy from 54.0\% to 100.0\%.

Although ConStruM uses fewer input tokens, it makes more calls and may cost
more than cached full-schema prompting, as observed on both
benchmarks (Table~\ref{tab:cost-comparison}). Full-schema prompting benefits
from discounted cached input, whereas ConStruM also pays for context
construction and additional generated output. Its principal benefit is higher
matching accuracy: \SurveyFullCombined{} combined mean accuracy, compared with
84.3\% for full-schema prompting (Table~\ref{tab:longitudinal-results}).

The demonstrated benefit also varies across tasks. HRS-B and PSID-B show
substantial gains on context-sensitive matching, while adding ConStruM to our
Matchmaker reconstruction changes accuracy little on MIMIC--OMOP and
Synthea--OMOP (Table~\ref{tab:mimic_top1}). These results support ConStruM's value
for contextual disambiguation, but do not establish comparable gains across
general schema-matching tasks.
\end{shepherding}

\section{Conclusion}

\begin{revision}
ConStruM turns schema structure into compact, query-specific evidence through
a context tree and similarity-group differentiation.  Across HRS-B and PSID-B,
it raises \shepadd{mean} combined accuracy from \shepadd{\SurveyNoContextCombined{}}
without context to \shepadd{\SurveyFullCombined{}} and ranks
first among all \shepadd{eleven} methods on both benchmarks.  It also outperforms a
full-schema LLM on both, showing that access to all metadata does not replace
selecting and organizing the evidence needed for a decision.  On
MIMIC--OMOP and Synthea--OMOP, attaching it to our Matchmaker reconstruction
produces no significant accuracy change.  The grouping and smaller-model
studies further show that its benefit is not confined to one grouping rule or
to GPT-5.4.  Together, these results present ConStruM as a modular way to make
schema context useful precisely where a correspondence depends on the
surrounding structure, while remaining compatible with a strong existing
matcher in the other settings tested.

Future work can enrich the context tree into a graph that captures relations
beyond hierarchy while retaining budgeted retrieval of decision-critical
evidence.  Structural signals from this graph, such as connectivity patterns
and the roles played by different fields, may provide further cues for
matching.
\end{revision}

\bibliographystyle{ACM-Reference-Format}
\bibliography{sample}

\clearpage
\appendix
\onecolumn
\input{appendix_content.tex}
\end{document}

%% file: generated/metrics.tex
\newcommand{\SurveyFullCombined}{95.4\%}
\newcommand{\SurveyNoContextCombined}{51.6\%}

%% file: authors.tex
\author{Houming Chen}
\orcid{0009-0000-7652-5485}
\affiliation{%
  \institution{University of Michigan}
  \city{Ann Arbor}
  \state{MI}
  \country{USA}}
\email{houmingc@umich.edu}

\author{Zhe Zhang}
\orcid{0009-0004-0548-0792}
\affiliation{%
  \institution{University of Michigan}
  \city{Ann Arbor}
  \state{MI}
  \country{USA}}
\email{zhezxz@umich.edu}

\author{H. V. Jagadish}
\orcid{0000-0003-0724-5214}
\affiliation{%
  \institution{University of Michigan}
  \city{Ann Arbor}
  \state{MI}
  \country{USA}}
\email{jag@umich.edu}

\renewcommand{\shortauthors}{Houming Chen, Zhe Zhang, and H. V. Jagadish}

%% file: figures/intro_time_example_tikz.tex
\begin{tikzpicture}[
  font=\small,
  attr/.style={draw, rounded corners, align=left, inner xsep=4pt, inner ysep=5pt, line width=0.6pt},
  query/.style={attr, fill=blue!10, line width=1.0pt},
  ctxnote/.style={draw, dashed, rounded corners, align=left, inner xsep=4pt, inner ysep=5pt, line width=0.6pt, fill=orange!10},
  cand/.style={attr, fill=gray!10},
]
  \node[cand, text width=0.44\columnwidth] (obs) {%
    \textbf{Candidate A:}\\
    \texttt{observation\_time}\\
    {\small time the observation occurred}%
  };

  \node[cand, text width=0.44\columnwidth, below=10pt of obs] (rec) {%
    \textbf{Candidate B:}\\
    \texttt{recorded\_time}\\
    {\small time the observation was recorded/entered}%
  };

  \coordinate (midcand) at ($(obs.west)!0.5!(rec.west)$);
  \node[query, anchor=east, text width=0.42\columnwidth] (chart) at ($(midcand)+(-0.45cm,0)$) {%
    \textbf{Query:}\\
    \texttt{CHARTTIME}\\
    {\small records the time at which an observation was made}%
  };

  \node[ctxnote, anchor=north east, text width=0.42\columnwidth] (ctx) at ($(chart.south east)+(0,-6pt)$) {%
    \textbf{Context (another column):} \texttt{STORETIME}\\
    {\small records the time at which an observation was manually input or manually validated}%
  };
  \draw[->, dashed, line width=0.7pt, color=orange!70!black] (ctx.north) -- (chart.south);

  \draw[->, line width=0.9pt, color=black!60] (chart.east) -- (obs.west);
  \draw[->, line width=0.9pt, color=black!60] (chart.east) -- (rec.west);
  \node[font=\footnotesize, fill=white, inner sep=1.2pt] at ($(chart.east)!0.55!(obs.west)$) {?};
  \node[font=\footnotesize, fill=white, inner sep=1.2pt] at ($(chart.east)!0.55!(rec.west)$) {?};

  \node[font=\small, align=left, text width=0.44\columnwidth, anchor=north west]
    at ($(rec.south west)+(0,-3pt)$) {%
      \texttt{STORETIME} is another column: it records when the observation was \emph{entered},
      so we know \texttt{CHARTTIME} is when it was \emph{made}; match \texttt{observation\_time}.%
    };
\end{tikzpicture}

%% file: figures/intro_strategy_comparison_float.tex
\begin{figure*}[t]
\centering
\Description{A three-panel diagram comparing LLM-based schema matching strategies: (a) local matching queries the LLM with a single source--target column pair; (b) global matching queries the LLM with all columns from both schemas, which is often impractical for wide schemas; (c) the proposed structure-guided approach derives a context tree and global similarity hypergraph, then uses only retrieved context and query-local confusable groups under a fixed budget.}
\begin{adjustbox}{max width=\linewidth}
\begin{tikzpicture}[
  font=\normalsize,
  colsrc/.style={circle, draw=blue!60, fill=blue!15, thick, minimum size=6mm},
  coltgt/.style={circle, draw=green!60, fill=green!15, thick, minimum size=6mm},
  llm/.style={rectangle, draw=black!60, rounded corners=2pt, fill=gray!10, thick, minimum width=27mm, minimum height=10mm, align=center},
  tinylabel/.style={midway, inner sep=1pt},
  plainlabel/.style={inner sep=1pt, align=center},
  dashedbox/.style={draw=black!30, dashed, rounded corners=2pt},
  arrow/.style={-{Latex[length=2mm]}, thick}
]

\def\titley{3.2}

\begin{scope}[xshift=-0.4cm]
\node[font=\small] at (-4.0,\titley) {\textbf{(a) Existing Method: Local}};
\node[colsrc] (s1) at (-5,2) {};
\node[coltgt] (t1) at (-3.4,2) {};
\node[plainlabel] at (-5,2.45) {source col};
\node[plainlabel] at (-3.4,2.45) {target col};
\node[llm]   (llmA) at (-4.2,0.6) {LLM};
\draw[arrow] (s1) -- node[tinylabel]{} (llmA);
\draw[arrow] (t1) -- (llmA);
\node[dashedbox, fit={(-6.0,-0.42) (-2.0,2.92)}] {};
\end{scope}

\begin{scope}[xshift=0.45cm]
\node[font=\small] at (0,\titley) {\textbf{(b) Existing Method: Global}};

\foreach \i in {0,...,11} {
  \pgfmathsetmacro{\x}{-1.8 + 0.3*mod(\i,4)} 
  \pgfmathsetmacro{\y}{2.3 - 0.3*floor(\i/4)} 
  \node[colsrc, minimum size=3mm] (gs\i) at (\x,\y) {};
}

\foreach \i in {0,...,11} {
  \pgfmathsetmacro{\x}{0.3 + 0.3*mod(\i,4)} 
  \pgfmathsetmacro{\y}{2.3 - 0.3*floor(\i/4)} 
  \node[coltgt, minimum size=3mm] (gt\i) at (\x,\y) {};
}

\node[llm] (llmB) at (0,0.6) {LLM};

\foreach \i in {0,...,11} {
  \draw[arrow, opacity=0.35] (gs\i) -- (llmB);
  \draw[arrow, opacity=0.35] (gt\i) -- (llmB);
}

\node[font=\small, align=center] at (0,-0.25) {entire schemas (\revadd{often impractical})};

\node[dashedbox, fit={(-2.0,-0.42) (2.0,2.92)}] {};
\end{scope}
\begin{scope}[xshift=1.15cm]
\node[font=\small] at (5.0,\titley) {\textbf{(c) Our Method: Structure-guided \& Context-aware}};

\node[colsrc, minimum size=5mm, fill=blue!50] (a) at (3.0,2.28) {}; 
\node[colsrc, minimum size=4.5mm, fill=blue!15] (b) at (2.56,2.00) {};
\node[colsrc, minimum size=4.5mm, fill=blue!15] (c) at (3.44,1.94) {};
\node[colsrc, minimum size=4.5mm, fill=blue!15] (d) at (3.46,2.60) {};
\node[colsrc, minimum size=4.5mm, fill=blue!15] (e) at (2.72,2.66) {};
\foreach \u/\v in {a/b,a/c,a/d,a/e,b/c,c/d} { \draw ( \u) -- (\v); }

\begin{scope}[on background layer]
  \node[draw=blue!50, fill=blue!8, rounded corners=3pt, fit={(2.42,1.80) (3.62,2.82)}] (srcctx) {};
\end{scope}

\node[coltgt, minimum size=5mm, fill=green!50] (p) at (5.0,2.28) {}; 
\node[coltgt, minimum size=4.5mm, fill=green!15] (q) at (5.46,2.02) {};
\node[coltgt, minimum size=4.5mm, fill=green!15] (r) at (5.84,2.56) {};
\node[coltgt, minimum size=4.5mm, fill=green!15] (u) at (5.10,1.76) {};
\node[coltgt, minimum size=4.5mm, fill=green!15] (v) at (5.84,1.86) {};
\foreach \u/\v in {p/q,p/r,q/v,r/v,u/q} { \draw ( \u) -- (\v); }

\begin{scope}[on background layer]
  \node[draw=green!50, fill=green!8, rounded corners=3pt, fit={(4.82,1.64) (6.00,2.74)}] (tgtctx) {};
\end{scope}

\node[llm] (llmC) at (4.2,0.28) {LLM};

\node[plainlabel, font=\small] at (2.78,1.25) {source\\context};
\node[plainlabel, font=\small] at (5.58,1.25) {target\\context};

\draw[arrow] (srcctx.south east) -- ([xshift=-5pt]llmC.north);
\draw[arrow] (tgtctx.south west) -- ([xshift=5pt]llmC.north);

\node[dashedbox, fit={(2.18,-0.42) (6.18,2.92)}] {};
\end{scope}

\end{tikzpicture}
\end{adjustbox}
\caption{Three approaches to LLM-based schema matching.
(a) \emph{Local}: the LLM reasons over an isolated source–target column pair, with little or no additional context.
(b) \emph{Global}: the LLM is provided with the full source and target schemas in a single prompt.
(c) \emph{Structure-guided (our approach)}: schema-derived structure is used to assemble, for each query column, a compact, query-specific set of contextual evidence that is supplied to the LLM for matching.}
\label{fig:local-global-graph}
\end{figure*}

%% file: figures/method_overview_flowchart_tikz.tex
\begin{tikzpicture}[
  font=\normalsize,
  box/.style={draw, rounded corners, align=center, inner xsep=4pt, inner ysep=3pt, line width=0.6pt, fill=gray!10},
  basebox/.style={box, fill=gray!08},
  conbox/.style={box, fill=blue!6, draw=black!55},
  conboxoffline/.style={box, fill=blue!6, draw=blue!55!black},
  conboxonline/.style={box, fill=green!6, draw=green!50!black},
  addonbox/.style={draw=black!35, dashed, rounded corners=2pt, inner sep=2pt},
  arrow/.style={-{Latex[length=2mm]}, line width=0.9pt, color=black!60, shorten <=1.2pt, shorten >=1.2pt},
  dashedarrow/.style={-{Latex[length=2mm]}, dashed, line width=0.9pt, color=black!60, shorten <=1.2pt, shorten >=1.2pt},
  label/.style={font=\footnotesize, color=black!65, inner sep=1pt},
]

  \coordinate (L) at (0,0);
  \coordinate (R) at (0.44\columnwidth,0);

  \node[basebox, text width=0.32\columnwidth, anchor=west] (bq) at ($(L)+(0,1.05)$) {Query\\{\small source col. \(s\)}};
  \node[basebox, text width=0.32\columnwidth, anchor=west, below=12pt of bq] (bup) {Upstream stages};
  \node[basebox, text width=0.32\columnwidth, anchor=west, below=12pt of bup] (bc0) {Shortlist\\{\small \(C_0 \subseteq T\)}};
  \node[basebox, text width=0.32\columnwidth, anchor=west, below=12pt of bc0] (bfinal) {Final decision\\{\small LLM: \ifdefined\deleted\deleted{\(t \in C_0\)}\else\shepadd{one target}\fi}};

  \draw[arrow] (bq.south) -- (bup.north);
  \draw[arrow] (bup.south) -- (bc0.north);
  \draw[arrow] (bc0.south) -- (bfinal.north);

  \node[align=center, anchor=south, font=\small\bfseries, text=black!90] at ($(bq.north)+(0,8pt)$)
    {General LLM-based matcher pipeline};

  \node[conboxoffline, text width=0.32\columnwidth, anchor=west] (schemas) at ($(R)+(0,1.05)$) {Schemas\\{\small source + target}};
  \node[conboxoffline, text width=0.32\columnwidth, anchor=west, below=6pt of schemas] (structures) {Structures\\{\small context tree + sim. hypergraph}};
  \node[conboxonline, text width=0.32\columnwidth, anchor=west, below=6pt of structures] (qin) {Inputs\\{\small \((s, C_0)\)}};
  \node[conboxonline, text width=0.32\columnwidth, anchor=west, below=6pt of qin] (pack) {Context pack};

  \node[addonbox, fit={(schemas) (structures) (qin) (pack)}] (addon) {};
  \node[anchor=south, font=\small\bfseries, text=black!90] at ($(addon.north)+(0,3pt)$) {ConStruM (add-on)};

  \begin{scope}[on background layer]
    \node[draw=blue!30, rounded corners=2pt, fill=blue!6, inner sep=1pt, fit={(schemas) (structures)}] (offline) {};
    \node[draw=green!30, rounded corners=2pt, fill=green!6, inner sep=1pt, fit={(qin) (pack)}] (online) {};
  \end{scope}
  \node[label, anchor=north west] at ($(offline.north west)+(2pt,-1pt)$) {\textbf{Offline}};
  \node[label, anchor=north west] at ($(online.north west)+(2pt,-1pt)$) {\textbf{Online}};

  \draw[arrow] (schemas.south) -- (structures.north);
  \path (qin.south) -- coordinate[pos=0.55] (qinpackmid) (pack.north);
  \draw[arrow] (qin.south) -- (pack.north);
  \coordinate (hookx) at ($(addon.east)+(6pt,0pt)$);
  \draw[dashedarrow, color=green!55!black, rounded corners=10pt]
    (structures.east) -- (hookx |- structures.east) |- (qinpackmid);

  \draw[arrow] (bc0.east) -- (qin.west);

  \draw[dashedarrow, color=green!55!black, line join=miter]
    (pack.west) -- (bfinal.east);
\end{tikzpicture}

%% file: appendix_content.tex
\begin{revision}
\section{Prompt and Construction Details}
\label{app:prompt-formats}

\paragraph{Context-pack serialization.}
The tree module renders each query or candidate as follows (abbreviated):
\begin{quote}\footnotesize\raggedright
\noindent
Column: [query column]\par
Description: [short description]\par
Path to root (summaries):\par
\hspace{0.8em}- [1] Query column leaf: [local summary]\par
\hspace{0.8em}- [2] Parent summary: [broader span summary]\par
\hspace{0.8em}- [3] Higher-level summary: [table/module-level scope]\par
\hspace{0.8em}- [\ldots] Dataset summary: [dataset-level scope cues]\par
Relation snippets (optional):\par
\hspace{0.8em}- [Field A qualifies the applicability of Field B]\par
\hspace{0.8em}- [Field C provides the unit/type for Field D]
\end{quote}

\paragraph{Differentiation-block serialization.}
Each non-singleton group receives a shared summary and one cue per field:
\begin{quote}\footnotesize\raggedright
\noindent
Differentiation for query (source-side group):\par
Summary: [commonality; differences in scope, stage, or applicability]\par
\hspace{0.8em}- query: [cue]\par
\hspace{0.8em}- sibling A: [cue]\par
Differentiation among candidates (Group \#1):\par
Summary: [common concept; differences in scope, units, or timeframe]\par
\hspace{0.8em}- cid 12: [candidate-specific cue]\par
\hspace{0.8em}- cid 37: [candidate-specific cue]
\end{quote}

\paragraph{Final decision prompt.}
The final call combines intrinsic metadata with the two evidence blocks:
\begin{quote}\footnotesize\ttfamily
\noindent
Query: name [\dots]; description [\dots]; context [\dots]\\
Source differentiation: [summary and per-field cues]\\
Candidates:\\
- cid 12: name [\dots]; description [\dots]; context [\dots]\\
- cid 37: name [\dots]; description [\dots]; context [\dots]\\
Candidate differentiation: [summary and per-candidate cues]
\end{quote}

\paragraph{Construction mechanics.}
The Stage-3 prompt asks for 2--10 conceptual groups and merges overlapping
themes; this is prompt guidance rather than a separate fan-out parameter.
Stage 4 avoids fragments shorter than five fields where possible and permits
at most three label switches per window unless one label is untenable.  The
stopping budget (B) is a target: \shepadd{recursion stops when a split makes no progress}.
Average-link cosine distance orders inter-table merges; an implementation may
cut the hierarchy at distance \(\delta\), although our multi-table experiments
retain the complete dendrogram.  Exact similarity-link construction is
\(O(n^2)\); an approximate index such as HNSW~\cite{malkov2018hnsw} can replace
it at larger scale.  A candidate introduced by expansion has no upstream
retrieval score; paired controls that use the same candidate roster serialize
that value as unavailable in both arms.

\section{Longitudinal Benchmarks: Per-Pair Results}
\label{app:hrs_b_pairwise}

\noindent
Table~\ref{tab:hrs_by_pair_full} reports correct/total and accuracy for every
wave pair under the five core methods in Table~1 \shepadd{from single
evaluations; Table~1 reports their repeated-run means};
misses and failures are included.
{
\scriptsize
\setlength{\tabcolsep}{2.4pt}
\begin{longtable}{@{}llrrrrr@{}}
  \caption{HRS-B and PSID-B per-pair forced-choice results
  \shepadd{from single evaluations}.}
  \label{tab:hrs_by_pair_full}\\
  \toprule
  \textbf{Benchmark} & \textbf{Year pair} & \textbf{Embed-1NN} & \textbf{No context} & \textbf{Tree only} &
  \textbf{Diff only} & \textbf{ConStruM} \\
  \midrule
\endfirsthead
  \caption[]{Longitudinal per-pair forced-choice results (continued).}\\
  \toprule
  \textbf{Benchmark} & \textbf{Year pair} & \textbf{Embed-1NN} & \textbf{No context} & \textbf{Tree only} &
  \textbf{Diff only} & \textbf{ConStruM} \\
  \midrule
\endhead
  \midrule
  \multicolumn{7}{r}{\scriptsize\emph{Continued on next page}} \\
\endfoot
  \bottomrule
\endlastfoot
    HRS-B & 2006$\rightarrow$2008 & 5/23 (21.7\%) & 19/23 (82.6\%) & 18/23 (78.3\%) & 23/23 (100.0\%) & 23/23 (100.0\%) \\
    & 2006$\rightarrow$2010 & 6/19 (31.6\%) & 19/19 (100.0\%) & 19/19 (100.0\%) & 19/19 (100.0\%) & 19/19 (100.0\%) \\
    & 2006$\rightarrow$2012 & 3/5 (60.0\%) & 3/5 (60.0\%) & 5/5 (100.0\%) & 5/5 (100.0\%) & 5/5 (100.0\%) \\
    & 2006$\rightarrow$2014 & 2/4 (50.0\%) & 4/4 (100.0\%) & 4/4 (100.0\%) & 4/4 (100.0\%) & 4/4 (100.0\%) \\
    & 2006$\rightarrow$2016 & 3/4 (75.0\%) & 4/4 (100.0\%) & 4/4 (100.0\%) & 4/4 (100.0\%) & 4/4 (100.0\%) \\
    & 2006$\rightarrow$2018 & 1/6 (16.7\%) & 0/6 (0.0\%) & 6/6 (100.0\%) & 6/6 (100.0\%) & 6/6 (100.0\%) \\
    & 2006$\rightarrow$2020 & 1/6 (16.7\%) & 1/6 (16.7\%) & 6/6 (100.0\%) & 6/6 (100.0\%) & 6/6 (100.0\%) \\
    & 2006$\rightarrow$2022 & 2/6 (33.3\%) & 0/6 (0.0\%) & 6/6 (100.0\%) & 6/6 (100.0\%) & 6/6 (100.0\%) \\
    & 2008$\rightarrow$2010 & 7/35 (20.0\%) & 29/35 (82.9\%) & 35/35 (100.0\%) & 34/35 (97.1\%) & 35/35 (100.0\%) \\
    & 2008$\rightarrow$2012 & 2/4 (50.0\%) & 2/4 (50.0\%) & 4/4 (100.0\%) & 4/4 (100.0\%) & 4/4 (100.0\%) \\
    & 2008$\rightarrow$2014 & 3/4 (75.0\%) & 2/4 (50.0\%) & 4/4 (100.0\%) & 4/4 (100.0\%) & 4/4 (100.0\%) \\
    & 2008$\rightarrow$2016 & 3/4 (75.0\%) & 3/4 (75.0\%) & 4/4 (100.0\%) & 4/4 (100.0\%) & 4/4 (100.0\%) \\
    & 2008$\rightarrow$2018 & 1/4 (25.0\%) & 1/4 (25.0\%) & 4/4 (100.0\%) & 4/4 (100.0\%) & 4/4 (100.0\%) \\
    & 2008$\rightarrow$2020 & 1/4 (25.0\%) & 0/4 (0.0\%) & 4/4 (100.0\%) & 4/4 (100.0\%) & 4/4 (100.0\%) \\
    & 2008$\rightarrow$2022 & 2/3 (66.7\%) & 1/3 (33.3\%) & 3/3 (100.0\%) & 3/3 (100.0\%) & 3/3 (100.0\%) \\
    & 2012$\rightarrow$2014 & 1/2 (50.0\%) & 2/2 (100.0\%) & 2/2 (100.0\%) & 2/2 (100.0\%) & 2/2 (100.0\%) \\
    & 2014$\rightarrow$2016 & 2/4 (50.0\%) & 4/4 (100.0\%) & 4/4 (100.0\%) & 4/4 (100.0\%) & 4/4 (100.0\%) \\
    & 2014$\rightarrow$2018 & 1/4 (25.0\%) & 0/4 (0.0\%) & 4/4 (100.0\%) & 4/4 (100.0\%) & 4/4 (100.0\%) \\
    & 2014$\rightarrow$2020 & 2/4 (50.0\%) & 1/4 (25.0\%) & 4/4 (100.0\%) & 4/4 (100.0\%) & 4/4 (100.0\%) \\
    & 2014$\rightarrow$2022 & 2/4 (50.0\%) & 0/4 (0.0\%) & 4/4 (100.0\%) & 4/4 (100.0\%) & 4/4 (100.0\%) \\
    & 2016$\rightarrow$2018 & 1/6 (16.7\%) & 1/6 (16.7\%) & 6/6 (100.0\%) & 6/6 (100.0\%) & 6/6 (100.0\%) \\
    & 2016$\rightarrow$2020 & 3/6 (50.0\%) & 1/6 (16.7\%) & 6/6 (100.0\%) & 4/6 (66.7\%) & 6/6 (100.0\%) \\
    & 2016$\rightarrow$2022 & 1/8 (12.5\%) & 2/8 (25.0\%) & 8/8 (100.0\%) & 8/8 (100.0\%) & 8/8 (100.0\%) \\
    & 2018$\rightarrow$2020 & 3/4 (75.0\%) & 3/4 (75.0\%) & 4/4 (100.0\%) & 4/4 (100.0\%) & 4/4 (100.0\%) \\
    & 2018$\rightarrow$2022 & 1/4 (25.0\%) & 0/4 (0.0\%) & 4/4 (100.0\%) & 4/4 (100.0\%) & 4/4 (100.0\%) \\
    & 2020$\rightarrow$2022 & 4/13 (30.8\%) & 2/13 (15.4\%) & 13/13 (100.0\%) & 11/13 (84.6\%) & 13/13 (100.0\%) \\
  \midrule
    & \textbf{HRS-B total} & \textbf{63/190 (33.2\%)} & \textbf{104/190 (54.7\%)} &
      \textbf{185/190 (97.4\%)} & \textbf{185/190 (97.4\%)} & \textbf{190/190 (100.0\%)} \\
  \midrule
    PSID-B & 2007$\rightarrow$2011 & 9/21 (42.9\%) & 15/21 (71.4\%) & 17/21 (81.0\%) & 20/21 (95.2\%) & 20/21 (95.2\%) \\
    & 2007$\rightarrow$2015 & 7/21 (33.3\%) & 7/21 (33.3\%) & 17/21 (81.0\%) & 17/21 (81.0\%) & 19/21 (90.5\%) \\
    & 2007$\rightarrow$2019 & 2/21 (9.5\%) & 6/21 (28.6\%) & 8/21 (38.1\%) & 13/21 (61.9\%) & 20/21 (95.2\%) \\
    & 2007$\rightarrow$2023 & 2/21 (9.5\%) & 2/21 (9.5\%) & 9/21 (42.9\%) & 14/21 (66.7\%) & 13/21 (61.9\%) \\
    & 2011$\rightarrow$2015 & 13/24 (54.2\%) & 18/24 (75.0\%) & 24/24 (100.0\%) & 23/24 (95.8\%) & 24/24 (100.0\%) \\
    & 2011$\rightarrow$2019 & 11/24 (45.8\%) & 14/24 (58.3\%) & 18/24 (75.0\%) & 23/24 (95.8\%) & 23/24 (95.8\%) \\
    & 2011$\rightarrow$2023 & 5/24 (20.8\%) & 8/24 (33.3\%) & 10/24 (41.7\%) & 20/24 (83.3\%) & 19/24 (79.2\%) \\
    & 2015$\rightarrow$2019 & 8/21 (38.1\%) & 14/21 (66.7\%) & 20/21 (95.2\%) & 19/21 (90.5\%) & 21/21 (100.0\%) \\
    & 2015$\rightarrow$2023 & 7/21 (33.3\%) & 10/21 (47.6\%) & 9/21 (42.9\%) & 15/21 (71.4\%) & 19/21 (90.5\%) \\
    & 2019$\rightarrow$2023 & 12/27 (44.4\%) & 17/27 (63.0\%) & 17/27 (63.0\%) & 23/27 (85.2\%) & 27/27 (100.0\%) \\
  \midrule
    & \textbf{PSID-B total} & \textbf{76/225 (33.8\%)} & \textbf{111/225 (49.3\%)} &
      \textbf{149/225 (66.2\%)} & \textbf{187/225 (83.1\%)} & \textbf{205/225 (91.1\%)} \\
\end{longtable}
}

\section{Parameter Settings and Practical Guidance}
\label{app:parameters}

\paragraph{Matchmaker add-on prompt.}
The final add-on prompt treats Matchmaker's confidence score as a strong prior
and changes the leading candidate only when ConStruM's context supplies a
specific reason, such as distinguishing same-named fields with different
table roles.

\paragraph{Broader-context baseline implementations.}
For ReMatch, we embed ten-field source and target neighborhoods with
\shepadd{the encoder specified in Section 7.2}, retrieve one target document, and apply the
recovered decision prompt to its ten fields.  For dense+BM25 RAG, we form
eight-field target chunks at stride four, embed their complete text, combine
dense and BM25 ranks by reciprocal-rank fusion ($k=60$), and expose the top
six chunks to the decision call.  For column-graph RAG, the 20
highest-cosine target fields seed a query-specific
column graph.  We add target--target similarity edges at cosine at least .95
and documentation-neighbor edges within radius two, retaining at most 160
nodes and 600 edges; node descriptions are truncated to 180 characters.  The
LLM sees this retrieved graph and must choose from the 20 seed candidates.
This operational baseline uses graph-organized retrieval, but does not run
GraphRAG's entity extraction, community detection, or community-summary
pipeline.  \shepadd{Two single evaluations under the identical protocol score}
116/190 (61.1\%) \shepadd{and} 119/190 (62.6\%); all 190 candidate lists
are identical, 176 predictions agree, and the paired
difference is not significant (exact McNemar $p=.375$).  For the full-schema
baseline, the prompt lists every source and target field identifier and
description, identifies the source query, and permits any target-schema
identifier as its answer; it contains no retrieved shortlist or ConStruM
evidence.  For
\textsc{LC-1/3}, source and target fields are each divided into
three consecutive, equal-row-count partitions, with
full field descriptions retained.  Each query therefore uses three proposal
calls---its source partition against each target partition---followed by one
arbitration call over the three proposed target IDs.

For the GPT-5.4-mini comparison, both arms \shepadd{use a compact decision
template and} apply the same label-blind repair:
an out-of-roster response receives one retry based on its prior response and
unchanged IDs, without labels, new evidence, retrieval, or candidates.
\shepadd{Across the three full-ConStruM runs, this triggers 26, 22, and 35 times;
4, 0, and 3 choices remain invalid afterward and count as incorrect.}
The no-context mini arm does not require the retry.

\noindent
Table~\ref{tab:parameter-defaults} summarizes the principal user-facing
controls and practical selection rules.  HRS-B and PSID-B use the same
context-stress configuration.  Lower-level parser constants are fixed
implementation settings recorded in the artifact rather than parameters
tuned to either benchmark.

\begin{table}[h]
  \centering
  \color{revisionred}
  \caption{Principal parameter values and practical guidance.}
  \label{tab:parameter-defaults}
  \scriptsize
  \setlength{\tabcolsep}{4pt}
  \begin{tabular}{@{}p{.18\textwidth}p{.18\textwidth}p{.18\textwidth}p{.40\textwidth}@{}}
    \toprule
    \textbf{Control} & \textbf{Context-stress value} & \textbf{General-matching value} &
      \textbf{Role and selection rule} \\
    \midrule
    Shortlist and decision cap & \(k=20\), cap 32 & upstream shortlist &
      Retains the top 20 and permits at most 12 context-expanded additions
      under a fixed prompt budget; the general add-on preserves its upstream
      candidate interface. \\
    Similarity threshold \(\tau\) & .95 & .90 &
      Defines confusable links.  Section~7.3 compares thresholds .90--.97 and
      five alternative grouping rules without using gold labels to set them. \\
    Local tree evidence & radius 6; last 6 path summaries & bounded path text &
      Controls the evidence shown at decision time and can be reduced when
      field descriptions are longer. \\
    Construction window \(W\) & 250 fields & wide parser inactive &
      Maximum fields in a wide-table construction call; reduce it to fit the
      construction model's input budget. \\
    Stopping budget \(B\) & 50 & wide parser inactive &
      Offline recursion target, not a hard leaf-size bound.  Match-time
      evidence is bounded independently. \\
    Inter-table cut \(\delta\) & inactive (one tree per wave) & inactive
      (complete hierarchy) & An optional hierarchy cut; none of the evaluated
      pipelines requires it. \\
    \bottomrule
  \end{tabular}
\end{table}
\end{revision}
\begin{shepherding}
\Needspace{10\baselineskip}
\section{Repeated-Run Statistical Analysis}
\label{app:repeat-statistics}

\paragraph{Per-benchmark intervals.}
For each benchmark, the accuracy interval is
$\bar a\pm t_{.975,2}s/\sqrt{3}$, where $\bar a$ and $s$ are the sample mean
and standard deviation of the three run accuracies.

\paragraph{Combined accuracy and uncertainty.}
Let $X_{mbr}$ be the number of correct decisions for method $m$ on benchmark
$b\in\{H,P\}$ in repetition $r$, with three repetitions per benchmark.
Writing $\bar X_{mb}$ and $s_{mb}^2$ for their sample mean and variance,
the merged-benchmark mean accuracy and its estimated variance are
\[
 \bar A_m=\frac{\bar X_{mH}+\bar X_{mP}}{415},\qquad
 v_{mb}=\frac{s_{mb}^2}{3\cdot415^2},\qquad V_m=v_{mH}+v_{mP}.
\]
Thus every decision counts equally; the separate variance components affect
uncertainty, not the relative contribution of either benchmark. For ConStruM
$c$ versus method $m$, the contrast is $D_m=\bar A_c-\bar A_m$, with estimated
variance $V_D=V_c+V_m$. Under independent matching repetitions, its approximate
Welch--Satterthwaite degrees of freedom are
\[
 \nu_D=\frac{V_D^2}
 {\displaystyle\sum_{j\in\{c,m\}}\sum_{b\in\{H,P\}}v_{jb}^2/(3-1)}.
\]
We test $D_m/\sqrt{V_D}$ against a two-sided $t_{\nu_D}$ distribution and
apply Holm correction to the ten combined comparisons. For each combined
accuracy interval, the same construction uses $V_m$ and
$\nu_m=V_m^2/\bigl[\sum_b v_{mb}^2/(3-1)\bigr]$, giving
$\bar A_m\pm t_{.975,\nu_m}\sqrt{V_m}$.
Intervals are pointwise, expressed as percentages and restricted to $[0,100]$.
This calculation uses separate benchmark variance estimates rather than
pairing run indices across surveys. Repeated questions do not become new
independent benchmark decisions. Zero observed variance produces a zero-width
descriptive interval, not evidence of deterministic behavior.

\paragraph{Benchmark-specific comparisons.}
Table~\ref{tab:survey-specific-tests} reports the secondary two-sided Welch
$t$-tests, with Holm correction across their 20 comparisons. They address a
different question from the merged-benchmark test: whether each method's
mean accuracy differs from ConStruM on an individual survey. All PSID-B
differences pass the corrected .05 threshold. The closer HRS-B differences
from full-schema prompting, LC-1/3, and differentiation only do not; the
HRS-B embedding and tree-only contrasts have no observed variance in either
series and their tests are not estimable. These results do not establish
equality. Both analyses describe matching variability on fixed benchmarks
with the reusable preprocessing specified in the paper, and rely on a
normal approximation with only three observations per series.

\begin{table}[h]
  \centering
  \color{shepherdred}
  \captionsetup{labelfont={color=shepherdred},textfont={color=shepherdred}}
  \caption{Secondary benchmark-specific comparisons with ConStruM.
  Entries are Holm-adjusted two-sided $p$-values; \emph{n.e.} denotes a test
  not estimable because both series have zero observed variance.}
  \label{tab:survey-specific-tests}
  \small
  \setlength{\tabcolsep}{12pt}
  \begin{tabular}{lcc}
    \toprule
    \textbf{Method} & \textbf{HRS-B} & \textbf{PSID-B} \\
    \midrule
Embed-1NN & \textit{n.e.} & 0.00193 \\
No-context LLM & 0.00153 & 0.00193 \\
ReMatch reimplementation & $4.28\times10^{-4}$ & 0.00193 \\
Matchmaker reconstruction & 0.00193 & $9.87\times10^{-4}$ \\
Dense+BM25 RAG & $1.77\times10^{-4}$ & $3.35\times10^{-5}$ \\
Column-graph RAG & 0.00193 & $7.20\times10^{-5}$ \\
Full-schema LLM & 0.278 & 0.0111 \\
LC-1/3 & 0.17 & 0.00248 \\
Tree only & \textit{n.e.} & 0.002 \\
Differentiation only & 0.17 & 0.0125 \\
    \bottomrule
  \end{tabular}
\end{table}
\end{shepherding}

%% file: sample.bib
@misc{sheetrit2024rematch,
  title={{ReMatch}: Retrieval Enhanced Schema Matching with {LLMs}},
  author={Sheetrit, Eitam and Brief, Menachem and Mishaeli, Moshik and Elisha, Oren},
  eprint={2403.01567},
  archivePrefix={arXiv},
  primaryClass={cs.DB},
  url={https://arxiv.org/abs/2403.01567},
  year={2024}
}

@inproceedings{parciak2024schema,
  title={Schema Matching with Large Language Models: an Experimental Study},
  author={Parciak, Marcel and Vandevoort, Brecht and Neven, Frank and Peeters, Liesbet M. and Vansummeren, Stijn},
  booktitle={Joint Workshops at the 50th International Conference on Very Large Data Bases ({VLDBW} 2024) --- {TaDA} 2024: 2nd International Workshop on Tabular Data Analysis},
  year={2024},
  url={https://vldb.org/workshops/2024/proceedings/TaDA/TaDA.8.pdf}
}

@inproceedings{seedat2024matchmaker,
  title={Bootstrapping Self-Improvement of Language Model Programs for Zero-Shot Schema Matching},
  author={Seedat, Nabeel and van der Schaar, Mihaela},
  booktitle={Proceedings of the 42nd International Conference on Machine Learning},
  series={Proceedings of Machine Learning Research},
  volume={267},
  pages={53791--53826},
  year={2025},
  publisher={PMLR},
  url={https://proceedings.mlr.press/v267/seedat25a.html}
}

@inproceedings{zhang-etal-2024-jellyfish,
  title={Jellyfish: Instruction-Tuning Local Large Language Models for Data Preprocessing},
  author={Zhang, Haochen and Dong, Yuyang and Xiao, Chuan and Oyamada, Masafumi},
  booktitle={Proceedings of the 2024 Conference on Empirical Methods in Natural Language Processing},
  pages={8754--8782},
  year={2024},
  publisher={Association for Computational Linguistics},
  doi={10.18653/v1/2024.emnlp-main.497}
}

@misc{gungor2025schemora,
  title={{Schemora}: Schema Matching via Multi-Stage Recommendation and Metadata Enrichment Using Off-the-Shelf {LLMs}},
  author={Gungor, Osman Erman and Paulsen, Derak and Kang, William},
  eprint={2507.14376},
  archivePrefix={arXiv},
  primaryClass={cs.DB},
  url={https://arxiv.org/abs/2507.14376},
  year={2025}
}

@article{zhang2024smutf,
  title={{SMUTF}: Schema Matching Using Generative Tags and Hybrid Features},
  author={Zhang, Yu and Mei, Di and Luo, Haozheng and Xu, Chenwei and Tsai, Richard Tzong-Han},
  journal={Information Systems},
  volume={133},
  pages={102570},
  year={2025},
  doi={10.1016/j.is.2025.102570}
}

@misc{jeon2025smog,
  title={Schema Matching on Graph: Iterative Graph Exploration for Efficient and Explainable Data Integration},
  author={Jeon, Mingyu and Suh, Jaeyoung and Cho, Suwan},
  eprint={2511.20285},
  archivePrefix={arXiv},
  primaryClass={cs.AI},
  url={https://arxiv.org/abs/2511.20285},
  year={2025}
}

@misc{ma2025knowledge,
  title={Knowledge Graph-based Retrieval-Augmented Generation for Schema Matching},
  author={Ma, Chuangtao and Chakrabarti, Sriom and Khan, Arijit and Moln{\'a}r, B{\'a}lint},
  eprint={2501.08686},
  archivePrefix={arXiv},
  primaryClass={cs.DB},
  url={https://arxiv.org/abs/2501.08686},
  year={2025}
}

@misc{xu2024kcmf,
  title={{KcMF}: A Knowledge-compliant Framework for Schema and Entity Matching with Fine-tuning-free {LLMs}},
  author={Xu, Yongqin and Li, Huan and Chen, Ke and Shou, Lidan},
  eprint={2410.12480},
  archivePrefix={arXiv},
  primaryClass={cs.CL},
  url={https://arxiv.org/abs/2410.12480},
  year={2024}
}

@inproceedings{zhang2021smat,
  title={{SMAT}: An Attention-Based Deep Learning Solution to the Automation of Schema Matching},
  author={Zhang, Jing and Shin, Bonggun and Choi, Jinho D. and Ho, Joyce C.},
  booktitle={Advances in Databases and Information Systems ({ADBIS} 2021)},
  series={Lecture Notes in Computer Science},
  volume={12843},
  pages={260--274},
  year={2021},
  publisher={Springer},
  address={Cham},
  doi={10.1007/978-3-030-82472-3_19}
}

@inproceedings{zhang2023plm,
  author    = {Yunjia Zhang and Avrilia Floratou and Joyce Cahoon and Subru Krishnan and Andreas C. M{\"u}ller and Dalitso Banda and Fotis Psallidas and Jignesh M. Patel},
  title     = {Schema Matching using Pre-Trained Language Models},
  booktitle = {2023 IEEE 39th International Conference on Data Engineering (ICDE)},
  pages     = {1558--1571},
  year      = {2023},
  publisher = {IEEE},
  doi       = {10.1109/ICDE55515.2023.00123}
}

@article{robertson2009bm25,
  author  = {Robertson, Stephen and Zaragoza, Hugo},
  title   = {The Probabilistic Relevance Framework: {BM25} and Beyond},
  journal = {Foundations and Trends in Information Retrieval},
  volume  = {3},
  number  = {4},
  pages   = {333--389},
  year    = {2009},
  doi     = {10.1561/1500000019}
}

@misc{HRS,
  title        = {Health and Retirement Study ({HRS})},
  author       = {{University of Michigan}},
  url          = {https://hrs.isr.umich.edu/},
  lastaccessed = {October 19, 2026},
  note         = {Produced and distributed by the University of Michigan with funding from the National Institute on Aging (grant number NIA U01AG009740)},
}

@article{malkov2018hnsw,
  title   = {Efficient and Robust Approximate Nearest Neighbor Search Using Hierarchical Navigable Small World Graphs},
  author  = {Malkov, Yu. A. and Yashunin, D. A.},
  journal = {IEEE Transactions on Pattern Analysis and Machine Intelligence},
  year    = {2020},
  volume  = {42},
  number  = {4},
  pages   = {824--836},
  doi     = {10.1109/TPAMI.2018.2889473},
  publisher = {IEEE}
}

@inproceedings{koutras2020rema,
  title={{REMA}: Graph Embeddings-based Relational Schema Matching},
  author={Koutras, Christos and Fragkoulis, Marios and Katsifodimos, Asterios and Lofi, Christoph},
  booktitle={Proceedings of the Workshops of the {EDBT/ICDT} 2020 Joint Conference},
  series={CEUR Workshop Proceedings},
  volume={2578},
  year={2020},
  publisher={CEUR-WS.org},
  url={https://ceur-ws.org/Vol-2578/SEAData5.pdf}
}

@article{tu2023unicorn,
  title={Unicorn: A Unified Multi-tasking Model for Supporting Matching Tasks in Data Integration},
  author={Tu, Jianhong and Fan, Ju and Tang, Nan and Wang, Peng and Li, Guoliang and Du, Xiaoyong and Jia, Xiaofeng and Gao, Song},
  journal={Proceedings of the ACM on Management of Data},
  volume={1},
  number={1},
  articleno={84},
  numpages={26},
  year={2023},
  doi={10.1145/3588938}
}

@article{Shraga2020ADnEVCS,
  title={{ADnEV}: Cross-Domain Schema Matching using Deep Similarity Matrix Adjustment and Evaluation},
  author={Roee Shraga and Avigdor Gal and Haggai Roitman},
  journal={Proceedings of the VLDB Endowment},
  year={2020},
  volume={13},
  number={9},
  pages={1401--1415},
  doi={10.14778/3397230.3397237}
}

@article{Li2020DeepEM,
  title={Deep Entity Matching with Pre-Trained Language Models},
  author={Yuliang Li and Jinfeng Li and Yoshihiko Suhara and AnHai Doan and Wang-Chiew Tan},
  journal={Proceedings of the VLDB Endowment},
  year={2020},
  volume={14},
  number={1},
  pages={50--60},
  doi={10.14778/3421424.3421431}
}

@inproceedings{melnik2002similarity,
  author    = {Sergey Melnik and Hector Garcia-Molina and Erhard Rahm},
  title     = {Similarity Flooding: A Versatile Graph Matching Algorithm and Its Application to Schema Matching},
  booktitle = {Proceedings of the 18th International Conference on Data Engineering},
  pages     = {117--128},
  year      = {2002},
  publisher = {IEEE},
  doi       = {10.1109/ICDE.2002.994702}
}

@inproceedings{Do2002COMAA,
  title={{COMA}---A System for Flexible Combination of Schema Matching Approaches},
  author={Hong-Hai Do and Erhard Rahm},
  booktitle={Proceedings of the 28th International Conference on Very Large Data Bases},
  pages={610--621},
  publisher={Morgan Kaufmann},
  year={2002},
  doi={10.1016/B978-155860869-6/50060-3}
}

@inproceedings{Gal2011UncertainSM,
  title={Uncertain Schema Matching: The Power of Not Knowing},
  author={Avigdor Gal},
  booktitle={Proceedings of the 20th ACM International Conference on Information and Knowledge Management},
  pages={2615--2616},
  publisher={ACM},
  year={2011},
  doi={10.1145/2063576.2064039}
}

@article{rahm2001survey,
  author  = {Erhard Rahm and Philip A. Bernstein},
  title   = {A survey of approaches to automatic schema matching},
  journal = {The {VLDB} Journal},
  volume  = {10},
  number  = {4},
  pages   = {334--350},
  year    = {2001},
  doi     = {10.1007/S007780100057}
}

@article{genossar2023battleship,
  author  = {Bar Genossar and Avigdor Gal and Roee Shraga},
  title   = {The Battleship Approach to the Low Resource Entity Matching Problem},
  journal = {Proceedings of the ACM on Management of Data},
  volume  = {1},
  number  = {4},
  articleno = {224},
  numpages = {25},
  year    = {2023},
  doi     = {10.1145/3626711}
}

@inproceedings{bohannon2006context,
  author    = {Philip Bohannon and Eiman Elnahrawy and Wenfei Fan and Michael Flaster},
  title     = {Putting Context into Schema Matching},
  booktitle = {Proceedings of the 32nd International Conference on Very Large Data Bases (VLDB)},
  pages     = {307--318},
  year      = {2006},
  publisher = {VLDB Endowment},
  url       = {https://www.vldb.org/conf/2006/p307-bohannon.pdf}
}

@inproceedings{sarthi2024raptor,
  title={{RAPTOR}: Recursive Abstractive Processing for Tree-Organized Retrieval},
  author={Sarthi, Parth and Abdullah, Salman and Tuli, Aditi and Khanna, Shubh and Goldie, Anna and Manning, Christopher D.},
  booktitle={The Twelfth International Conference on Learning Representations},
  year={2024},
  url={https://proceedings.iclr.cc/paper_files/paper/2024/hash/8a2acd174940dbca361a6398a4f9df91-Abstract-Conference.html}
}

@article{liu2024lost,
  title={Lost in the Middle: How Language Models Use Long Contexts},
  author={Liu, Nelson F. and Lin, Kevin and Hewitt, John and Paranjape, Ashwin and Bevilacqua, Michele and Petroni, Fabio and Liang, Percy},
  journal={Transactions of the Association for Computational Linguistics},
  volume={12},
  pages={157--173},
  year={2024},
  doi={10.1162/tacl_a_00638}
}

@inproceedings{bai2024longbench,
  title={{LongBench}: A Bilingual, Multitask Benchmark for Long Context Understanding},
  author={Bai, Yushi and Lv, Xin and Zhang, Jiajie and Lyu, Hongchang and Tang, Jiankai and Huang, Zhidian and Du, Zhengxiao and Liu, Xiao and Zeng, Aohan and Hou, Lei and Dong, Yuxiao and Tang, Jie and Li, Juanzi},
  booktitle={Proceedings of the 62nd Annual Meeting of the Association for Computational Linguistics (Volume 1: Long Papers)},
  pages={3119--3137},
  year={2024},
  publisher={Association for Computational Linguistics},
  doi={10.18653/v1/2024.acl-long.172}
}

@inproceedings{liu2024gram,
  title={{GRAM}: Generative Retrieval Augmented Matching of Data Schemas in the Context of Data Security},
  author={Liu, Xuanqing and Wang, Runhui and Song, Yang and Kong, Luyang},
  booktitle={Proceedings of the 30th ACM SIGKDD Conference on Knowledge Discovery and Data Mining},
  pages={5476--5486},
  year={2024},
  publisher={ACM},
  doi={10.1145/3637528.3671602}
}

@misc{edge2024graphrag,
  title={From Local to Global: A Graph {RAG} Approach to Query-Focused Summarization},
  author={Edge, Darren and Trinh, Ha and Cheng, Newman and Bradley, Joshua and Chao, Alex and Mody, Apurva and Truitt, Steven and Metropolitansky, Dasha and Ness, Robert Osazuwa and Larson, Jonathan},
  year={2024},
  eprint={2404.16130},
  archivePrefix={arXiv},
  primaryClass={cs.CL},
  url={https://arxiv.org/abs/2404.16130}
}

@inproceedings{guo2025lightrag,
  title = {{LightRAG}: Simple and Fast Retrieval-Augmented Generation},
  author = {Guo, Zirui and Xia, Lianghao and Yu, Yanhua and Ao, Tu and Huang, Chao},
  booktitle = {Findings of the Association for Computational Linguistics: EMNLP 2025},
  pages = {10746--10761},
  address = {Suzhou, China},
  month = {November},
  year = {2025},
  publisher = {Association for Computational Linguistics},
  doi = {10.18653/v1/2025.findings-emnlp.568}
}

@incollection{wang2025llmatch,
  title={{LLMATCH}: A Unified Schema Matching Framework with Large Language Models},
  author={Wang, Sha and Li, Yuchen and Xiao, Hanhua and Dai, Bing Tian and Lee, Roy Ka-Wei and Dong, Yanfei and Deng, Lambert},
  booktitle={Web and Big Data},
  series={Lecture Notes in Computer Science},
  volume={16116},
  pages={343--356},
  year={2026},
  publisher={Springer},
  address={Singapore},
  doi={10.1007/978-981-95-5722-6_37}
}

@article{liu2025magneto,
  title={Magneto: Combining Small and Large Language Models for Schema Matching},
  author={Liu, Yurong and Pena, Eduardo H. M. and Santos, A{\'e}cio and Wu, Eden and Freire, Juliana},
  journal={Proceedings of the VLDB Endowment},
  volume={18},
  number={8},
  pages={2681--2694},
  year={2025},
  doi={10.14778/3742728.3742757}
}

@article{johnson2016mimiciii,
  author  = {Johnson, Alistair E. W. and Pollard, Tom J. and Shen, Lu and Lehman, Li-Wei H. and Feng, Mengling and Ghassemi, Mohammad and Moody, Benjamin and Szolovits, Peter and Celi, Leo Anthony and Mark, Roger G.},
  title   = {{MIMIC-III}, a freely accessible critical care database},
  journal = {Scientific Data},
  year    = {2016},
  volume  = {3},
  pages   = {160035},
  doi     = {10.1038/sdata.2016.35}
}

@article{overhage2012omopcdm,
  author  = {Overhage, J. Marc and Ryan, Patrick B. and Reich, Christian G. and Hartzema, Abraham G. and Stang, Paul E.},
  title   = {Validation of a common data model for active safety surveillance research},
  journal = {Journal of the American Medical Informatics Association},
  year    = {2012},
  volume  = {19},
  number  = {1},
  pages   = {54--60},
  doi     = {10.1136/amiajnl-2011-000376},
  pmid    = {22037893},
  pmcid   = {PMC3240764},
  url     = {https://www.ncbi.nlm.nih.gov/pmc/articles/PMC3240764/}
}

@article{doan2005semanticintegration,
  author  = {Doan, AnHai and Halevy, Alon Y.},
  title   = {Semantic Integration Research in the Database Community: A Brief Survey},
  journal = {AI Magazine},
  year    = {2005},
  volume  = {26},
  number  = {1},
  pages   = {83--94},
  doi     = {10.1609/aimag.v26i1.1801},
  url     = {https://ojs.aaai.org/aimagazine/index.php/aimagazine/article/view/1801}
}

@book{bellahsene2011schemamatchingmapping,
  title     = {Schema Matching and Mapping},
  editor    = {Bellahsene, Zohra and Bonifati, Angela and Rahm, Erhard},
  publisher = {Springer},
  series    = {Data-Centric Systems and Applications},
  year      = {2011},
  doi       = {10.1007/978-3-642-16518-4},
  url       = {https://link.springer.com/book/10.1007/978-3-642-16518-4}
}

@article{deng2021turl,
  author  = {Deng, Xiang and Sun, Huan and Lees, Alyssa and Wu, You and Yu, Cong},
  title   = {{TURL}: Table Understanding through Representation Learning},
  journal = {Proceedings of the {VLDB} Endowment},
  year    = {2021},
  volume  = {14},
  number  = {3},
  pages   = {307--319},
  doi     = {10.14778/3430915.3430921}
}

@article{fan2023starmie,
  author  = {Fan, Grace and Wang, Jin and Li, Yuliang and Zhang, Dan and Miller, Ren{\'e}e J.},
  title   = {Semantics-aware Dataset Discovery from Data Lakes with Contextualized Column-based Representation Learning},
  journal = {Proceedings of the {VLDB} Endowment},
  year    = {2023},
  volume  = {16},
  number  = {7},
  pages   = {1726--1739},
  doi     = {10.14778/3587136.3587146}
}

@misc{psid,
  author       = {{Survey Research Center}},
  title        = {Panel Study of Income Dynamics, Public Use Dataset},
  howpublished = {Institute for Social Research, University of Michigan,
                  Ann Arbor, Michigan},
  note         = {Produced and distributed by the Survey Research Center},
  year         = {2026},
  url          = {https://psidonline.isr.umich.edu/},
  lastaccessed = {August 9, 2026}
}

@inproceedings{ester1996dbscan,
  author    = {Martin Ester and Hans-Peter Kriegel and J{\"o}rg Sander and Xiaowei Xu},
  title     = {A Density-Based Algorithm for Discovering Clusters in Large Spatial Databases with Noise},
  booktitle = {Proceedings of the Second International Conference on Knowledge Discovery and Data Mining},
  pages     = {226--231},
  year      = {1996},
  publisher = {AAAI Press},
  url       = {https://aaai.org/papers/kdd96-037-a-density-based-algorithm-for-discovering-clusters-in-large-spatial-databases-with-noise/}
}

@article{murtagh2012algorithms,
  author  = {Fionn Murtagh and Pedro Contreras},
  title   = {Algorithms for Hierarchical Clustering: An Overview},
  journal = {WIREs Data Mining and Knowledge Discovery},
  volume  = {2},
  number  = {1},
  pages   = {86--97},
  year    = {2012},
  doi     = {10.1002/widm.53}
}

@article{welch1947generalization,
  author  = {Welch, B. L.},
  title   = {The Generalization of {`Student's'} Problem when Several Different Population Variances are Involved},
  journal = {Biometrika},
  volume  = {34},
  number  = {1--2},
  pages   = {28--35},
  year    = {1947},
  doi     = {10.1093/biomet/34.1-2.28}
}

@article{holm1979simple,
  author  = {Holm, Sture},
  title   = {A Simple Sequentially Rejective Multiple Test Procedure},
  journal = {Scandinavian Journal of Statistics},
  volume  = {6},
  number  = {2},
  pages   = {65--70},
  year    = {1979},
  url     = {https://www.jstor.org/stable/4615733}
}

@article{satterthwaite1946approximate,
  author  = {Satterthwaite, F. E.},
  title   = {An Approximate Distribution of Estimates of Variance Components},
  journal = {Biometrics Bulletin},
  volume  = {2},
  number  = {6},
  pages   = {110--114},
  year    = {1946},
  doi     = {10.2307/3002019}
}
